\documentclass[pdflatex,sn-mathphys-num]{sn-jnl}% Math and Physical Sciences Numbered Reference Style
\usepackage{graphicx}%
\usepackage{multirow}%
\usepackage{amsmath,amssymb,amsfonts}%
\usepackage{amsthm}%
\usepackage{mathrsfs}%
\usepackage[title]{appendix}%
\usepackage{xcolor}%
\usepackage{textcomp}%
\usepackage{manyfoot}%
\usepackage{booktabs}%
\usepackage{enumitem}
\usepackage{algorithm}%
\usepackage{algorithmicx}%
\usepackage{algpseudocode}%
\usepackage{listings}%
\usepackage{longtable,booktabs}   % Necessário para usar tabelas que quebram páginas
\usepackage{array}       % Necessário para definir colunas personalizadas (P{...})
\usepackage{ragged2e}    % Para usar \RaggedRight corretamente dentro das colunas
\usepackage{pgfplots}
\pgfplotsset{compat=1.18} % ou 1.17 se preferir
\usepgfplotslibrary{statistics} % NECESSÁRIO para histogramas
\usepgfplotslibrary{groupplots} % <- NECESSÁRIO para \begin{groupplot}
\usepackage{xcolor} % opcional, mas útil se quiser ajustar cores
\usepackage{subcaption}

\pgfplotsset{
  every axis/.style={
    tick label style={font=\scriptsize},
    label style={font=\scriptsize},
    title style={font=\footnotesize},
    grid=both,
    grid style={line width=0.2pt, draw=black!10},
    major grid style={draw=black!20}
  },
  scatterstyle/.style={
    only marks,
    mark size=2.2pt,
    line width=0.5pt
  }
}

\theoremstyle{thmstyleone}%
\theoremstyle{thmstyletwo}%

\theoremstyle{thmstylethree}%

\begin{document}

\title[Article Title]{Evaluating the Reliability of Multiple Large Language Models in Risk Assessment: A CIS Controls‑Based Approach}


\author[1]{\fnm{Gustavo Roberto} \sur{Pinto}}\email{gus@ufu.br}

\author[1]{\fnm{Arthur do Prado} \sur{Labaki}}\email{arthur.labaki@ufu.br}
\equalcont{These authors contributed equally to this work.}

\author*[1]{\fnm{Rodrigo Sanches} \sur{Miani}}\email{miani@ufu.br}
\equalcont{These authors contributed equally to this work.}

\affil*[1]{\orgdiv{Faculty of Computing}, \orgname{Federal University of Uberlândia}, \orgaddress{\street{Av. João Naves de Ávila, 2121}, \city{Uberlândia}, \postcode{38408-100}, \state{Minas Gerais}, \country{Brazil}}}

%%\affil[2]{\orgdiv{Department}, \orgname{Organization}, \orgaddress{\street{Street}, \city{City}, \postcode{10587}, \state{State}, \country{Country}}}

%%\affil[3]{\orgdiv{Department}, \orgname{Organization}, \orgaddress{\street{Street}, \city{City}, \postcode{610101}, \state{State}, \country{Country}}}


\abstract{Proper implementation of technical and administrative controls reinforces an organization's cybersecurity posture and business resilience, reduces risks, and enhances governance, ultimately elevating business maturity. The dynamics of the technological landscape and emerging threats negatively affect the most diverse companies, regardless of their size. This, associated with a global gap in the cybersecurity workforce, imposes enormous challenges and the need for a profound change in how companies respond to threats. Generative Artificial Intelligence from large language models has become an influential tool across various companies, emerging as a viable option to help address those challenges while partially addressing the shortage of skilled labor. Although large language models can help in this scenario, there may be risks, such as generating unreliable or 'hallucinated' content, which could lead people and companies to make bad decisions. Our study proposes integrating human experts into the validation process as a crucial step toward ensuring the proper implementation of technical and administrative controls. Furthermore, we sought to identify how large language models perform in assessing cybersecurity risk scenarios compared to human experts, highlighting the importance of integrating humans and machines in the cybersecurity risk assessment process. Using a questionnaire with risk scenarios, we analyzed responses from 50 human experts. We compared their responses with those of five popular large language models to determine whether it is possible to use only large language models for cybersecurity risk assessment. The results reveal that the large language models consistently underestimated cybersecurity risks compared to human experts, reinforcing the need for human oversight and suggesting that LLMs should be used as complementary tools rather than standalone assessors.}




\maketitle

\section{Introduction}\label{sec1}

Organizations today face an increasingly complex threat landscape where the proper implementation of technical and administrative controls directly impacts cybersecurity posture, business resilience, and overall governance. However, the rapid evolution of the technology landscape and the rise of sophisticated threats—such as ransomware attacks, supply chain compromises, zero-day vulnerabilities, and advanced persistent threats (APTs)—exacerbated by geopolitical tensions, have put immense pressure on security teams, regardless of company size. The 20th edition of the Global Risk Report for 2025, produced exclusively by the World Economic Forum \cite{in01}, lists the main risks for the next 2 and 10 years. Cyber Espionage and Warfare appear as the fifth most severe risk for the next two years, while Adverse Outcomes of AI Technologies appear as the sixth most significant risk for the next 10 years. Compounding these challenges with a worldwide shortage of qualified cybersecurity professionals prompts organizations to explore innovative methods for strengthening their defenses \cite{in02}. The Cybersecurity Workforce Study, produced by (ISC)2 in 2024, estimates a gap of around 4.7 million cybersecurity professionals, an increase of 19\%.

In response, generative artificial intelligence—particularly large language models—has emerged as a promising tool \cite{in03}. By automating specific tasks and accelerating risk assessments, AI can help partially address the workforce gap. However, this potential comes with pitfalls: large language models sometimes generate unreliable or "hallucinated" outputs, leading to misguided decisions if not properly scrutinized. A human expert-in-the-loop approach becomes critical to mitigate such risks, enabling skilled practitioners to validate and refine the AI's recommendations.

Although recent studies explore the use of generative AI across various cybersecurity domains, there remains a lack of research evaluating large language models' ability to perform cybersecurity risk assessments independently \cite{in05, in06, in07, in08, in09, in10}. This paper addresses that gap by comparing the performance of an LLM to human participants across structured scenarios, aiming to assess its potential to approximate expert judgment.

We investigate how five LLMs (ChatGPT 5, DeepSeek V3.1, Llama 4 Scout, Claude Opus 4.1, and Gemini 2.5 PRO) assign risk levels to cybersecurity scenarios aligned with recognized frameworks, such as the CIS CONTROLS \cite{in11}. By comparing Large Language Model (LLM) risk ratings to those assigned by human cybersecurity professionals and measuring how much they diverge across multiple controls, we aim to clarify the extent to which LLMs can be effectively used in conjunction with human expertise for cybersecurity risk assessment. Therefore, this work investigates the performance of multiple LLMs, compares it to the performance of human cybersecurity professionals, and analyzes the results, aiming to answer the following research questions (RQs): 

\begin{enumerate}
    \item To what extent can LLMs be trusted to conduct cybersecurity risk assessments without the involvement of human experts? 
    \item  \textit{How LLMs used in this study compare to human cybersecurity professionals in assessing risk across standardized scenarios }
    \begin{itemize}
        \item \textbf{Senior and Specialist Professionals:} \textit{How does the LLMs performance align with or differ from that of senior and specialist cybersecurity professionals who possess advanced expertise and experience?}
        \item \textbf{Other Professionals (Non-Senior/Specialist):} \textit{How do LLMs compare to professionals at other levels (e.g., junior, mid-level, interns) regarding risk assessment accuracy and consistency?}
    \end{itemize}
    \item Do different LLMs produce significantly different risk ratings for the same cybersecurity scenarios?
\end{enumerate}

This study offers three main contributions to the field of cybersecurity risk assessment. First, we provide an extensive comparative analysis between human cybersecurity professionals and multiple state-of-the-art large language models in the context of structured risk-assessment scenarios aligned with CIS Controls, offering empirical evidence on how LLMs interpret and rate cybersecurity risks. Second, we demonstrate that LLMs systematically underestimate risk severity across different scenario categories and professional seniority levels, revealing concrete limitations that challenge their use as standalone decision-makers in cybersecurity governance. Finally, we propose and validate a human-expert-in-the-loop approach as an essential mechanism to mitigate hallucinations and unreliable outputs, positioning LLMs as complementary tools that can support—but not replace—expert judgment in risk-assessment workflows. Collectively, these contributions address a critical gap in the literature and provide practical insights for organizations considering the integration of LLMs into cybersecurity processes.

The remainder of this article is organized as follows. Section 2 (Background) introduces key concepts referenced and discussed throughout the paper. Section 3 (Literature Review) presents the main works related to the topic and reviews relevant studies that have informed the development of this research. Section 4 (Methodology) outlines the structured approach adopted in the study. Section 5 (Results and Discussion) presents the data and analyzes the findings. Finally, Section 6 (Conclusion and Future Work) summarizes the main conclusions and outlines future work.

\section{Background}\label{sec2}

This section explores two closely connected topics: Generative Artificial Intelligence (Generative AI) and Large Language Models (LLMs), alongside Security Controls and Frameworks. The first subsection examines the development, functionalities, and applications of Generative AI and LLMs, emphasizing their influence across industries while addressing their advantages and ethical challenges. The second subsection focuses on the significance of security controls and frameworks, underscoring their critical role in protecting organizations from cyber threats.

\subsection{Generative AI and Large Language Models}

Generative Artificial Intelligence (Generative AI) and Large Language Models (LLMs), such as OpenAI's ChatGPT, Google's Gemini, Meta's Llama, and Anthropic's Claude, have gained prominence in recent years for their ability to assist professionals across various fields and tasks. These models leverage vast neural networks and deep learning techniques to process and generate text, code, and even images naturally, mimicking complex human cognitive functions such as problem-solving, creativity, and decision-making \cite{bg01, bg02}.

Generative AI encompasses various artificial intelligence methods that create original, meaningful content by learning from extensive datasets. A notable milestone in this field was the development of Generative Adversarial Networks (GANs), which introduced a groundbreaking adversarial training approach. GANs utilize two competing neural networks—a generator and a discriminator—to produce highly realistic synthetic images, audio, and videos \cite{bg03}. Despite this achievement, the introduction of the Transformer architecture in 2017 significantly reshaped Generative AI by enabling the emergence of sophisticated Large Language Models. The Transformers' attention mechanism and self-attention allowed these models to process and understand extensive textual contexts, producing coherent, contextually accurate text outputs, thus dramatically expanding the capabilities and applications of Generative AI beyond visual content generation \cite{bg02, bg04}.

Large Language Models represent a powerful application of Generative AI, specifically focused on language processing and generation. These models are trained on massive, diverse datasets, including books, research articles, websites, and structured knowledge bases, enabling them to learn and generalize linguistic patterns effectively. Such robust training empowers LLMs to execute complex tasks, including text translation, summarization, sentiment analysis, and even source code generation in various programming languages \cite{in03, bg01}.

Furthermore, the creative potential of Large Language Models has significantly impacted content generation. These models can effectively write narratives, draft scripts, generate compelling advertising copy, and even assist with on-demand brainstorming for innovative ideas, streamlining and enhancing creative processes across industries from entertainment to marketing. This transformative potential highlights their role as technological tools and as essential components in creative fields \cite{bg02, bg05}.

However, Generative AI and Large Language Models pose several critical challenges alongside their evident advantages, such as ethical concerns, including the potential to generate misinformation, reinforce existing biases, or unintentionally propagate stereotypes inherent in their training data. These models inherit biases from the vast volumes of data they are trained on, necessitating rigorous evaluation protocols and ethical guidelines to mitigate potential harm and ensure the responsible deployment of AI technologies \cite{in06, bg05}.

Large Language Models have emerged as powerful tools in cybersecurity, significantly influencing defensive and offensive capabilities. On the defensive side, they assist in analyzing massive volumes of security logs and network traffic data, efficiently identifying anomalies, predicting attack patterns, and automating routine security operations. Recent approaches demonstrate that combining LLMs with traditional static code analysis significantly improves vulnerability detection, underscoring the promising role of neuro-symbolic AI methodologies in enhancing cybersecurity resilience \cite{in07}. On the other hand, the potential misuse of LLMs by cybercriminals has raised significant alarms within the security community, as adversaries can exploit the technology to craft highly persuasive phishing emails or automate the discovery and exploitation of software vulnerabilities on a large scale \cite{in08}.

These opportunities and threats underline the critical importance of careful evaluation and strategic planning when deploying LLMs in sensitive fields such as cybersecurity. The establishment of comprehensive frameworks for evaluating AI capabilities and ongoing research into the practical and ethical implications of Generative AI are essential steps toward balanced and responsible technology integration \cite{in06, in08}.

\subsection{Security Controls and Frameworks}

Security controls and frameworks are crucial in protecting organizations against increasingly sophisticated cyber threats. These structured tools provide a consistent foundation for identifying, mitigating, and managing risks across various sectors. In addition to offering clear guidelines, frameworks help align organizational practices with global security best practices \cite{bg11}.

One widely recognized framework is the Center for Internet Security (CIS) Critical Security Controls, which provides a practical, prioritized approach to securing an organization's assets. The CIS consists of specific controls categorized into priority levels, allowing businesses to scale their security measures based on size, resources, and operational complexity \cite{in11}. Implementing CIS controls helps mitigate the most common threats and is a starting point for organizations seeking to structure their security strategy.

Another key pillar is the NIST Cybersecurity Framework (CSF), developed by the National Institute of Standards and Technology. This framework provides a flexible and adaptable approach based on five core functions: Identify, Protect, Detect, Respond, and Recover. These functions form a continuous improvement cycle, enabling organizations to proactively address cyber threats and optimize their security processes. Due to its broad applicability and governmental support, the NIST CSF is widely adopted by both private companies and public institutions \cite{bg13}.

The ISO/IEC 27001, published by the International Organization for Standardization, outlines the requirements for implementing an Information Security Management System (ISMS). This framework is particularly valued for its focus on cybersecurity governance, covering everything from access control policies to regular audits. ISO 27001 certification has become a global standard, demonstrating an organization's commitment to information protection and building trust with customers and partners \cite{bg14}.

Beyond being operational tools, these frameworks have strategic significance for cybersecurity. They serve as references for regulatory compliance and foster cross-sector collaboration by sharing best practices and lessons learned. By adopting security frameworks, organizations can consistently enhance their security posture, thereby strengthening the global cybersecurity network \cite{in09}.

\section{Related Work}

Research on the intersection of cybersecurity and artificial intelligence spans a broad spectrum of approaches, ranging from LLM-based benchmarks and specialized evaluation frameworks to practical tools and methodologies for organizational risk management. To situate our contribution within this landscape, we structured this section into two complementary parts. First, we review studies that explore the use of Large Language Models (LLMs) and other AI-driven techniques for cybersecurity tasks, including benchmarks, specialized models, and evaluation methods that assess LLM capabilities across technical domains. Second, we examine research focused on cybersecurity risk analysis, encompassing both traditional frameworks and tools—often geared toward SMEs—and more recent AI-supported approaches. This organization allows us to contrast AI-centric methods with governance-, risk-, and compliance-oriented solutions, highlighting the gap our work addresses.

Unlike prior studies, which typically evaluate model performance on factual knowledge, Q\&A benchmarks, or specialized detection tasks, our work adopts a structure closer to Governance, Risk, and Compliance (GRC) audits. We use simulated scenarios with structured question-and-answer sets grounded in CIS Controls and compare numerical risk assessments (0–10) produced by two groups: (a) 50 human cybersecurity professionals, categorized by experience, and (b) multiple state-of-the-art LLMs. This design enables the analysis of statistical correlation, the identification of systematic risk underestimation by LLMs, and the evaluation of the feasibility of hybrid human–AI approaches. By focusing on risk scoring rather than purely diagnostic, factual, or classification tasks, our study provides a distinct perspective within the broader research landscape.

\subsection{Use of Large Language Models (LLMs) in Cybersecurity}

The authors of \cite{rw02} propose SECURE, a comprehensive benchmark for evaluating LLMs in cybersecurity tasks, with particular emphasis on industrial control systems (ICS). The benchmark spans six task sets involving vulnerability interpretation, exploitation reasoning, interpreter abuse, and technical advisory generation, drawing on sources such as CWE, CVE, MITRE ATT\&CK, and CISA. The evaluation of seven LLMs—including GPT-4, Gemini, and Llama—revealed substantial performance gaps, underscoring the importance of domain-specific and realistic benchmarks.

In \cite{rw03}, the authors introduce MEQA, a meta-evaluation framework designed to assess the quality of Q\&A benchmarks applied to LLMs in cybersecurity contexts. MEQA defines eight criteria and 44 subcriteria, covering aspects such as validity, reliability, evaluator design, robustness, and reproducibility. By applying this framework with both human evaluators and LLMs, the study highlights methodological weaknesses in existing benchmarks, including limited representativeness and low statistical stability.

The work in \cite{rw04} presents CyberMetric, a benchmark for assessing the factual and technical knowledge of LLMs across nine cybersecurity domains using multiple-choice questions validated by experts. Through Retrieval-Augmented Generation (RAG), CyberMetric offers a consistent structure for comparing model and human performance under closed-book conditions focused on technical knowledge.

A different perspective is presented in \cite{rw05}, where the authors propose the Cybersecurity Evaluation Tool (CET), a maturity assessment instrument tailored for SMEs and grounded in the NIST framework. CET includes a questionnaire with 35 critical controls, reporting mechanisms, and recommended actions, enabling small organizations to interpret and apply cybersecurity guidance despite limited expertise.

\subsection{Risk Analysis in Cybersecurity}

In \cite{rw06}, the authors propose CyberGen, a methodology that combines LLMs and knowledge graphs (KGs) to generate validated cybersecurity educational content automatically. Using zero-shot, few-shot, and ontology-driven prompting strategies, CyberGen produces approximately 4,000 expert-verified Q\&A pairs in the CyberQ dataset, supporting scalable training and instructional design in cybersecurity.

The authors of \cite{rw07} introduce SecurityBERT, a lightweight model optimized for cyber-threat detection in IoT and IIoT environments. Based on a BERT architecture enhanced with a privacy-preserving feature learning encoding (PPFLE) approach, the model achieved 98.2\% accuracy on the Edge-IIoT dataset and millisecond-level inference times, making it suitable for resource-constrained environments.

In \cite{rw08}, the authors present CyberPal.AI, a family of cybersecurity-focused LLMs trained on the SecKnowledge dataset, which comprises over 400,000 instructions generated via expert curation and synthetic techniques. The accompanying SecKnowledge-Eval benchmark, comprising 15 evaluation sets, demonstrates the models’ robust performance across complex reasoning tasks in threat detection, explanation, and analysis.

The qualitative study reported in \cite{in03} investigates barriers faced by small and medium-sized enterprises (SMEs) in adopting cybersecurity practices. Through interviews with European SMEs, the authors identify misconceptions about attack likelihood, informal organizational structures, unclear accountability, and skepticism toward specialized providers. The results highlight the importance of aligning technical recommendations with cultural and operational realities.

The authors of \cite{bg11} propose a COBIT-based self-assessment tool designed to guide SMEs toward ISO/IEC 27001 compliance. Structured around 14 security domains and evaluated using maturity levels, the tool is validated through three case studies demonstrating its feasibility for organizations with minimal cybersecurity expertise.

In \cite{rw09}, the authors introduce Yacraf, a quantitative cybersecurity risk assessment model combining threat modeling, attack trees, and a structured metamodel. Compatible with FAIR methodology, Yacraf provides stronger visual and analytical support for decision-making and is validated through real-world corporate deployments.

Finally, \cite{rw10} proposes a risk-assessment approach specifically tailored to SMEs, grounded in Protection Motivation Theory (PMT) and Self-Determination Theory (SDT). The GEIGER score derived from the model offers intuitive risk interpretation and personalized countermeasures while preserving user privacy.

\subsection{Comparative Analysis of Related Work}

Overall, the related work highlights a broad spectrum of initiatives that leverage LLMs and other AI-driven methods for cybersecurity, ranging from benchmarking efforts for model evaluation—such as SECURE \cite{rw02}, MEQA \cite{rw03}, CyberMetric \cite{rw04}, and SecKnowledge-Eval \cite{rw08}—to practical tools and organizational frameworks designed for SMEs, including CET \cite{rw05}, COBIT-to-ISO self-assessment approaches \cite{bg11}, and GEIGER-based risk scoring models \cite{rw10}. Research also examines specialized models such as SecurityBERT \cite{rw07} and CyberPal.AI \cite{rw08}, as well as innovative methodologies like CyberGen \cite{rw06}, and Yacraf \cite{rw09}, which address tasks ranging from threat detection and structured risk analysis to the generation of educational content. Collectively, these studies underscore the growing role of LLMs and structured frameworks in advancing cybersecurity while also exposing persistent challenges, including model limitations, the influence of human factors, and the need for scalable, domain-specific solutions. In this context, the present research distinguishes itself by evaluating risk assessments grounded in CIS Controls, systematically comparing LLM-generated ratings with those of human experts, and examining the viability of hybrid approaches that integrate human judgment with AI-driven support.

Table~\ref{tab:compar-trab} presents a comparative summary of the main studies discussed above. The analyzed works vary widely in scope, from benchmarks and specialized AI models to practical assessment tools for SMEs and structured risk-analysis frameworks. This synthesis helps contextualize the novelty of the present work, particularly its contribution in evaluating LLM-based risk assessments derived from CIS Controls and systematically comparing them with human expert judgments.

\setlength{\LTcapwidth}{\textwidth}
\setlength{\LTleft}{0pt}
\setlength{\LTright}{0pt}
\newcolumntype{P}[1]{>{\RaggedRight\arraybackslash}p{#1}}

\begingroup
\setlength{\tabcolsep}{2pt}       % tighter columns
\renewcommand{\arraystretch}{1.35}% taller rows
{\fontsize{7pt}{8pt}\selectfont
\begin{longtable}{|P{2.2cm}|P{1.5cm}|P{1.8cm}|P{1.9cm}|P{1.3cm}|P{1.3cm}|P{2.2cm}|}
\caption{Comparison between related works and the proposed study}\label{tab:compar-trab}\\
\hline
\textbf{Ref.} & \textbf{Category} & \textbf{Domain} & \textbf{Output/Task} & \textbf{LLM vs hum.} & \textbf{Risk} & \textbf{Bases/Refs.}\\
\hline
\endfirsthead
\hline
\textbf{Ref.} & \textbf{Category} & \textbf{Domain} & \textbf{Output/Task} & \textbf{LLM vs hum.} & \textbf{Risk} & \textbf{Bases/Refs.}\\
\hline
\endhead
\hline
\multicolumn{7}{r}{\textit{Continued on next page}}\\
\endfoot
\hline
\endlastfoot

\cite{rw02} SECURE & Benchmark & ICS & Multi-task (reasoning) & \textbf{Yes} & No & CWE, CVE, MITRE ATT\&CK, CISA \\ \hline
\cite{rw03} MEQA & Meta-evaluation & LLM benchmarks & Criteria (8+44) & \textbf{Yes} & No & — \\ \hline
\cite{rw04} CyberMetric & Benchmark & Tech. knowledge & MCQ (closed-book) & \textbf{Yes} & No & RAG, human validation \\ \hline
\cite{rw05} CET (SMEs) & Tool & SMEs / NIST & Questionnaire + recommendations & No & \textbf{Yes} & NIST CSF \\ \hline
\cite{rw06} CyberGen/CyberQ & Method / Dataset & Education & Q\&A generation & No & No & KG / Ontology \\ \hline
\cite{rw07} SecurityBERT & Model & IoT / IIoT & Threat detection & No & No & — \\ \hline
\cite{rw08} CyberPal.AI & LLMs & Threats / analysis & Instructions + tasks & No & No & MITRE, CVE, CAPEC, Sigma, SIEM \\ \hline
\cite{in03} Qualitative study (SMEs) & Study & SMEs / adoption & Factors and barriers & N/A & No & — \\ \hline
\cite{bg11} COBIT$\rightarrow$ISO & Tool & SMEs / governance & Maturity by domains & No & Partial & COBIT 4.1, ISO/IEC 27001 \\ \hline
\cite{rw09} Yacraf & Framework & Quantitative risk & Attack tree + calculation & No & \textbf{Yes} & FAIR-like / metamodel \\ \hline
\cite{rw10} GEIGER (SMEs) & Method / Tool & SMEs & Risk score + actions & No & \textbf{Yes} & PMT, SDT, CERT, GEIGER \\ \hline
This work & Comparative study & GRC / audit & Risk score (0--10) & \textbf{Yes} & \textbf{Yes} & CIS Controls \\ \hline

\end{longtable}
}
\endgroup

% ---------------------------------------------------------------------------

%%%%%%%%%%%%%%%%

\section{Methodology} 

\begin{figure*}
\centerline{\includegraphics[width=4.5in]{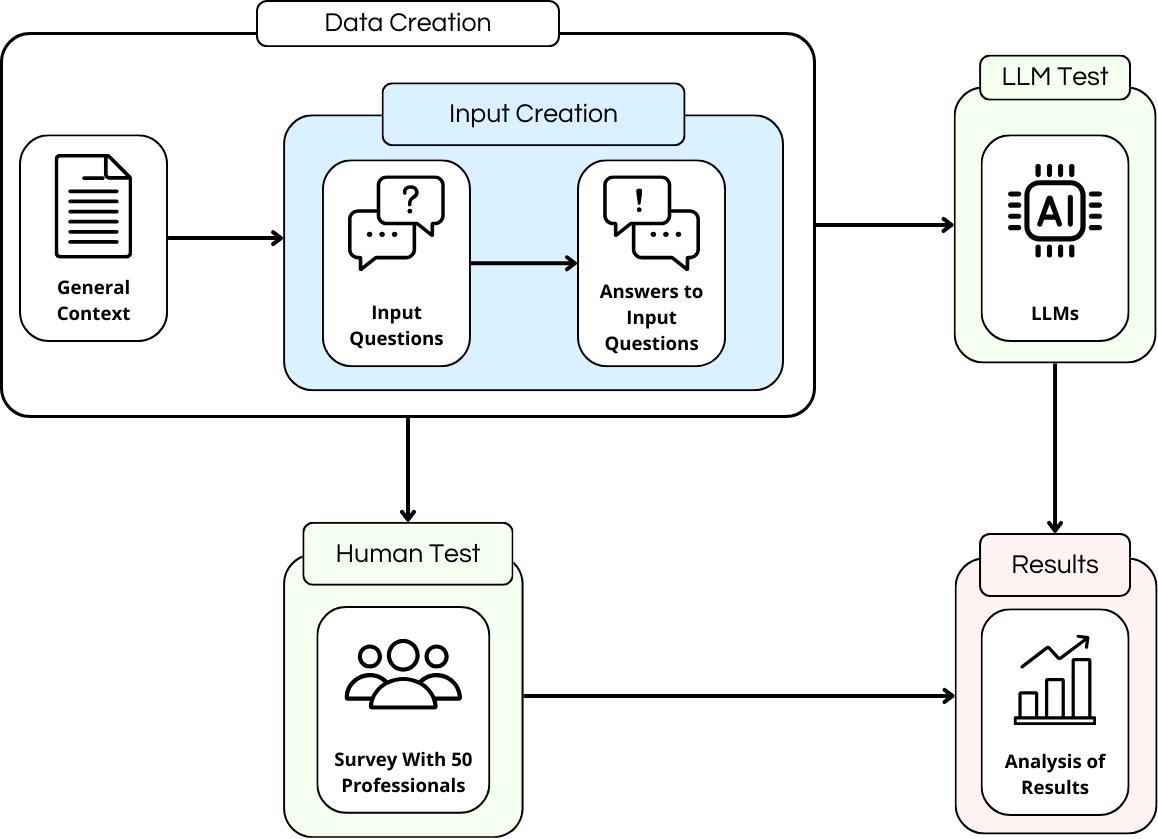}}
\caption{Flowchart for cybersecurity risk assessment using LLMs and human evaluation.
\label{fig_meth01}}
\end{figure*}

Our main goal is to evaluate the effectiveness of a Large Language Model in assessing cybersecurity risks and compare its performance with that of human professionals at different levels of expertise. To achieve this, we designed a structured methodology divided into four main phases, as illustrated in Figure \ref{fig_meth01}. These phases follow a sequential approach to ensure a systematic and rigorous evaluation of the risk assessment process. Next, we describe each phase in detail.

\begin{enumerate}[label=\Alph*)]
\item \textbf{Data Creation:} This phase involves the development of contextualized input data to guide the risk assessment process. We defined a general context, specifying a cybersecurity scenario in which an information security analyst must assess the security risk of a partner company. Based on this, we created a set of structured questions and corresponding responses to simulate real-world assessments, which we called Input Creation.

    \begin{itemize}
        \item \textbf{Input Creation:} In this stage, we developed a set of questions based on the CIS Framework, ensuring that a corresponding question represented each control. We analyzed various real-world assessments to enhance the realism and quality of the responses. Drawing on these insights, we crafted responses aligned with a simulated scenario representing a mid-sized technology company.
    \end{itemize}

\item \textbf{LLM Test:} In this step, the input data were tested using the selected language models. We fed the models structured responses to assess their ability to assess cybersecurity risks. Different input variations were tested, including extreme and mixed cases, to evaluate the model’s consistency in risk assessment.

\item \textbf{Human Test:} We conducted a complementary human evaluation by recruiting participants with diverse backgrounds in cybersecurity and IT. These participants assessed security risks using the same contextual scenario used in the LLM evaluation. We prepared a structured questionnaire to collect their assessments and demographic and professional background information.

\item \textbf{Results Analysis:} Finally, we conducted a statistical analysis of the responses from both the LLMs and the human participants. Our study examined the distribution of risk classifications, the correlation between human and AI evaluations, and applied statistical significance tests to identify potential trends or discrepancies in risk perception.

\end{enumerate}

%% Infra do texto:
% A) Data Creation
%   - Input Creation
% B) LLM
% C) Human
% D) Result

\subsection{Data Creation}

When working with LLMs for text generation, it is essential to develop a general context that establishes the current environment and outlines what needs to be done, while exposing as many details as possible \cite{bg01}. This general context allows the LLM to understand its real purpose, reducing the chances of misinterpretation. Additionally, context is essential for helping individuals understand a specific subject \cite{bg02}. Thus, the developed general context is:

\textit{"You are an information security analyst working for a medium-sized company. This company handles sensitive customer information, from personally identifiable data to strategic records. Currently, the company is concerned about the information security posture adopted by its partners.}

\textit{These partners play an essential role in the supply chain, providing services that support critical business processes and functions. They require access to network resources, including internal and external applications, data, and IT services. To assess the risks associated with these partners, the company uses the CIS (Critical Security Controls) framework, which enables the identification of potential risks before service delivery begins.}

\textit{Currently, one of the partners is being evaluated regarding the cybersecurity risks associated with their infrastructure, processes, systems, applications, and other aspects. The necessary data for this evaluation are presented below, including customized questions aligned with the CIS standards and the partner's answers.}

\textit{Your objective, as an information security analyst, is to determine the security risk associated with each partner based on your expertise, assigning a rating on a scale from 0 (low risk) to 10 (high risk), to define the level of risk each partner represents to the company."}

\subsubsection*{Input Creation}

The input creation process was based on the details described in the general context. It consisted of two main elements: the questions posed by the security analyst and the responses provided by the evaluated partner. These inputs were carefully designed to simulate a realistic cybersecurity risk assessment scenario, ensuring consistency and authenticity in the evaluation process.

We used the CIS v8 framework as a reference for developing the questions. We selected a security topic for each of the 18 controls that best represented the essential practices associated with that control. Based on these topics, we formulated clear, objective questions that reflect the type of inquiry a security analyst might direct to the evaluated partner. These questions were designed to comprehensively capture the key aspects of each control, ensuring that the information gathered would be relevant and valuable for the risk assessment.

The responses were designed to reflect how business partners typically answer security assessments. To achieve this, we analyzed previously conducted security reports and evaluations to identify patterns in expected responses. We observed that partners generally have a positive outlook on their security practices, while acknowledging certain limitations and challenges. Additionally, their responses tend to be detailed, often including more information than explicitly requested, which helps minimize interpretive ambiguity.

The final set of inputs was structured to include the formulated questions and the simulated responses. The responses followed a realistic pattern, predominantly positive but incorporating relevant negative aspects to ensure balance. This pattern was defined based on the analysis of real-world assessments, reflecting how business partners typically respond to such evaluations. This approach allows us to assess whether the language model can accurately identify and weigh risks, just as a human analyst would. The complete input can be visualized below:

\begin{enumerate}
  \item \textbf{Theme:} Hardware Asset Inventory 
  \textbf{Question:} Does the company have a detailed inventory of corporate assets?
  \textbf{Answer:} The company maintains a comprehensive and detailed inventory of assets. This inventory covers the vast majority of the organization's assets, with approximately 98% accounted for and tracked. The inventory is continuously updated through automated systems, ensuring that new assets are captured and existing ones are updated in real time as changes occur.

  \item \textbf{Theme:} Software Asset Inventory and Control 
  \textbf{Question:} Does the company have a detailed inventory of software assets, systems, and applications?
  \textbf{Answer:} The organization has a well-established and effectively maintained software inventory. Currently, approximately 85\% of the organization's software is inventoried, ensuring comprehensive control over software assets. This inventory is regularly updated through manual processes, ensuring that new installations and updates are reflected in the system.
  
  \item \textbf{Theme:} Information Management 
  \textbf{Question:} Is there a formal data management process in place?
  \textbf{Answer:} A formal data management process is in place within the organization. Data is classified by sensitivity. The data management process is audited every six months, and access permissions are regularly reviewed.
  
  \item \textbf{Theme:} Secure Configuration Process (hardening, security baselines, etc.) \textbf{Question:} Is there a secure configuration process for all systems?
  \textbf{Answer:} A secure configuration process is in place for all systems, based on the CIS (Center for Internet Security) framework. Configurations are standardized across the organization and are reviewed and updated quarterly.

  \item \textbf{Theme:} User Account Inventories 
  \textbf{Question:} Is there an inventory of all user accounts across the systems listed in the application catalog?
  \textbf{Answer:} An inventory of all system user accounts in the application catalog is maintained. This inventory is frequently and automatically updated. Accounts are centrally managed, and unused accounts are promptly deactivated.

  \item \textbf{Theme:} Access Provisioning in Systems and Applications 
  \textbf{Question:} Is there a formal process for granting access?
  \textbf{Answer:} A formal access-granting process is in place. In this process, requests are approved based on the principle of least privilege, and access is periodically reviewed to ensure it is still necessary. Additionally, granted accesses are role- or function-based.

  \item \textbf{Theme:} Vulnerability Management 
  \textbf{Question:} Is there a vulnerability management process in place for all assets (endpoints, servers, network devices, applications, systems, etc.)?
  \textbf{Answer:} A vulnerability management process exists for all assets, including endpoints, servers, network devices, applications, and systems. Vulnerabilities are identified and prioritized for remediation. Vulnerability scans are conducted monthly, and management reports are regularly reviewed.

  \item \textbf{Theme:} Log and Activity Record Management 
  \textbf{Question:} Is there a formal audit log management process in place?
  \textbf{Answer:} A formal audit log management process is in place, with automatic log collection from all systems and secure storage in centralized and encrypted repositories. Dedicated teams regularly review logs using automated tools to detect suspicious activities. According to the policy, audit logs are retained for 12 months in a secure environment, with access restricted to authorized users.

  \item \textbf{Theme:} Use of Browsers and Email Clients 
  \textbf{Question:} Are all browsers and email clients fully supported, and are only authorized versions implemented?
  \textbf{Answer:} All browsers and email clients in use are fully supported, with regular updates and ongoing vendor support. Security settings for browsers and email clients are centrally enforced and monitored to ensure compliance. The list of supported versions is reviewed and updated quarterly to ensure the latest and most secure versions are in use.

  \item \textbf{Theme:} Antimalware 
  \textbf{Question:} Is antimalware software deployed on all endpoints?
  \textbf{Answer:} Antimalware software is deployed on all organizational endpoints, with regular malware scans conducted. The effectiveness of the antimalware software is tested quarterly. The anti-exploit feature is enabled, and endpoints are centrally managed through a dedicated security platform.

  \item \textbf{Theme:} Data and Information Recovery 
  \textbf{Question:} Is there a data recovery process in place?
  \textbf{Answer:} A data recovery process is in place, with backups performed regularly and stored securely. Critical data can be restored within 8 hours in the event of a loss. Recovery logs are maintained and periodically reviewed to ensure the integrity of the process.

  \item \textbf{Theme:} Keeping Network Infrastructure Updated 
  \textbf{Question:} Is the network infrastructure regularly updated?
  \textbf{Answer:} The network infrastructure is regularly updated, with firmware and software patches promptly applied. All configuration changes to network devices are tracked and documented. The network infrastructure is audited for compliance semiannually.

  \item \textbf{Theme:} Centralization of Information Security Events 
  \textbf{Question:} Are all information security and related events centralized?
  \textbf{Answer:} All security event alerts are centralized on a dedicated platform, enabling real-time monitoring of the entire infrastructure. All critical events are sent to this central system, where they are analyzed and addressed by specialized teams.

  \item \textbf{Theme:} Awareness Program 
  \textbf{Question:} Is there currently a program to raise awareness or train employees on cybersecurity topics?
  \textbf{Answer:} An active security awareness program is in place, and all employees must complete security awareness training. Training is conducted annually.

  \item \textbf{Theme:} Third-Party, Partner, and Supplier Management 
  \textbf{Question:} How is third-party, partner, and supplier management currently conducted regarding the mapping and analysis of information security risks?
  \textbf{Answer:} An inventory of all service providers is maintained and regularly updated. Service providers are periodically reviewed for security compliance, and their security practices are evaluated annually.

  \item \textbf{Theme:} Secure Software Development 
  \textbf{Question:} Are secure software development practices implemented?
  \textbf{Answer:} A Secure Application Development Process is established, with security practices embedded in all stages of the development lifecycle. Development teams follow secure coding standards and conduct regular security reviews, such as vulnerability testing and code audits. Security updates and patches are applied continuously to ensure application protection.

  \item \textbf{Theme:} Incident Response 
  \textbf{Question:} Is there an incident response process in place?
  \textbf{Answer:} An Incident Response Process is in place, with a designated team responsible for handling incidents. Incident handling training is conducted regularly, and all incidents are documented and reviewed.

  \item \textbf{Theme:} Penetration Testing and Attack Simulations 
  \textbf{Question:} Are penetration tests, security tests, or attack simulations conducted?
  \textbf{Answer:} A Penetration Testing Program is in place, with regular tests conducted on critical systems and network infrastructure. Specialized teams carry out these tests to identify and exploit vulnerabilities, ensuring corrective actions are promptly taken.
\end{enumerate}

\subsection{LLMs Test}
Before applying the language model to the primary dataset developed for this study, we conducted a preliminary analysis to verify whether the selected LLMs could generate coherent and contextually grounded responses in the context of risk assessments. To this end, we used the same questions from the main dataset but created an experimental set of simplified responses, as illustrated in Figure \ref{fig_meth02}. We designed four distinct scenarios, with each question receiving four variations of responses:

\begin{figure}
\centerline{\includegraphics[width=4.8in]{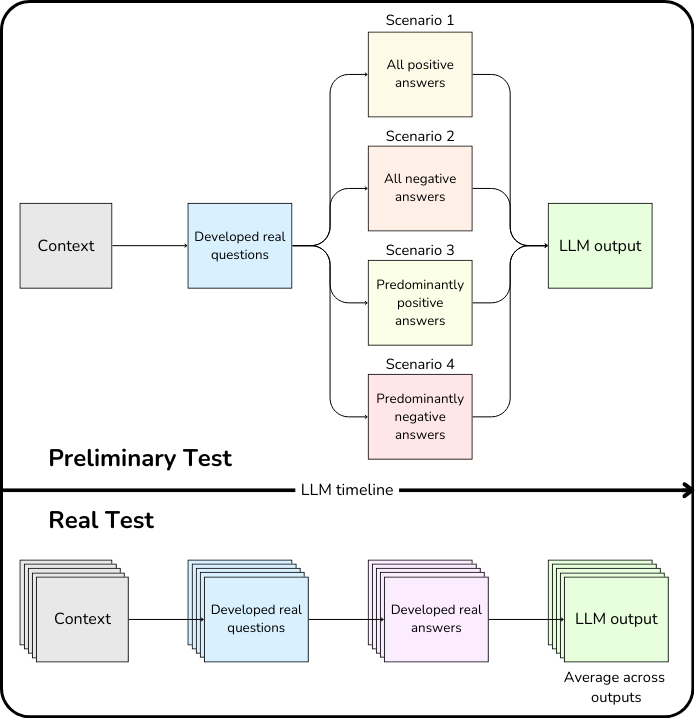}}
\caption{Flowchart illustrating the difference between the preliminary test and the real test.
\label{fig_meth02}}
\end{figure}

\begin{enumerate}
    \item All answers as "Yes" for every question.
    \item All answers as "No" for every question.
    \item Mixed answers, but predominantly "Yes".
    \item Mixed answers, but predominantly "No".
\end{enumerate}

This test allowed us to evaluate how the LLMs interpreted extreme and moderate scenarios and to verify whether they could recognize patterns and provide justified assessments rather than superficial or arbitrary responses. The results demonstrated that all models accurately interpreted response variations and assigned coherent risk classifications without resorting to extreme or inconsistent evaluations. Consequently, this confirmed that all five models (ChatGPT 5, DeepSeek V3.1, Llama 4 Scout, Claude Opus 4.1, and Gemini 2.5 PRO) were suitable for application in this study.

Following this preliminary phase, we conducted the main evaluation using the original dataset, explicitly designed to enable comparisons between the LLMs and human cybersecurity experts. The model analyzed each provided response, assigning a risk level and detailed justifications for every evaluation. Additionally, the models were asked to give an overall risk classification for the entire scenario. To ensure consistency, we repeated the tests across different conversations and accounts, verifying the stability of the results. In all executions, the LLMs demonstrated consistency in their classifications and justifications, highlighting their analytical capability within the context of cybersecurity risk assessment.

To minimize potential variations in responses and ensure greater statistical robustness, each evaluation was performed five times in independent conversations for each model. Accordingly, the values used in the final analysis are the average of the five tests conducted per model, providing a more balanced and reliable view of the LLMs' performance.

\subsection{Human Test}
We recruited 50 participants through our social media networks and contacted individuals from universities, research institutions, and technology companies to volunteer. All participants were informed, through the online form used for data collection, that their anonymized responses would be used solely for academic research purposes. By voluntarily completing and submitting the form, participants provided their informed consent for the use of their anonymized data in this study. According to Brazilian regulations for research involving human subjects—specifically the Resolution CNS nº 510/2016, which governs ethical oversight for studies in the humanities and social sciences—research activities that gather anonymous opinion data without any possibility of identifying participants are exempt from review by an Institutional Review Board (IRB)/ethics committee. Therefore, no ethics committee approval was required for this study. 

Similar to the LLMs Test, this Human Test phase involved recruiting people to complete a form in the same context provided to the LLM, and to classify the input according to the level of risk it posed to the company. To achieve this, we developed a Google Forms survey containing questions about the participants' field of expertise, profile type, years of experience, professional level, risk level classifications for each input, and an optional field to justify the assigned risk.

The participants ranged from individuals who had recently entered the field (whether in security or IT in general) to professionals with over 10 years of proven experience and internationally recognized certifications. Additionally, to ensure impartial responses in the survey, volunteers were asked to complete the questionnaire without external assistance. Finally, we conducted a statistical analysis to confirm that all obtained responses were valid and consistent.

\subsection{Results}

The survey data and LLM responses were analyzed using descriptive statistics to identify response patterns, focusing on the mean, dispersion, and distribution of risk classifications assigned by participants and the five distinct LLMs. To deepen the understanding of differences between groups, we segmented the participants based on their professional level and compared them across two main groups:

\begin{itemize}
    \item Group 0: All Participants
    \item Group 1: Highly specialized professionals (Senior and Specialists)
    \item Group 2: Less specialized professionals (Entry-level and Intermediate)
\end{itemize}
 
In parallel, we compared the classifications assigned by the LLMs with the averages of human experts to identify discrepancies and possible trends in the models' evaluations. Additionally, we compared the LLMs to determine which performed better in the context of information security risk assessment. This comparison provides valuable insights into the relative strengths and weaknesses of each LLM, highlighting which approaches may be more effective for practical cybersecurity applications.

To verify the significance of the observed differences, we applied statistical tests to determine whether the differences between groups were statistically significant. Furthermore, we analyzed the linear correlation between participants' responses and the models' evaluations to determine whether both followed a similar pattern in risk perception. We also examined the distribution of assigned scores to understand the range and frequency of values provided by humans and by the LLMs. This approach helps identify differences and determine whether they occur systematically or randomly.

To ensure the validity of the results, we adopted several measures to minimize biases and ensure sample representativeness. The participant selection included professionals from different levels and fields within Information Security, providing a diverse range of perspectives. In addition, the questionnaire was administered anonymously and without external influence, reducing biases in individual judgments. Finally, the robustness of the analysis was reinforced by rigorous statistical techniques, which enabled a reliable interpretation of the results and contributed to a better understanding of differences in risk perception between humans and AIs.

\section{Results and discussions}

Next, we present the results of the data analysis conducted throughout our study, along with our responses to the proposed research questions. The results are organized as follows: in Subsection~\ref{results:demographics}, we describe the characteristics of the 50 participants, including their professional level, years of experience, primary area of activity, and professional profile. We discuss and address research questions RQ1 and RQ2 in Subsections~\ref{subsection_rq1} and~\ref{subsection_rq2}, respectively. Finally, we present our discussion and the study's limitations in Subsection~\ref{res:dis}.

\subsection{Demographics}
\label{results:demographics}
We engaged 50 participants for our survey through social networks, private discussion groups, and direct contact with individuals from universities, research institutions, and companies. The survey was designed using Google Forms and included questions about participants' years of experience, professional level, professional profile, and area of expertise. To ensure unbiased responses, volunteers—ranging from beginners to experts—were asked to complete the questionnaire without external assistance. 

To better understand the characteristics of our survey participants, we conducted a statistical analysis of their demographic and professional attributes. The dataset includes information on years of experience, professional level, job profile, and area of expertise. This analysis provides insights into the distribution of knowledge among respondents, highlighting variations in experience levels and industry representation.

Tables \ref{tab:res_01}, \ref{tab:res_02}, \ref{tab:res_03}, and \ref{tab:res_04} summarize the distribution of participants across the main demographic and professional categories. These breakdowns offer a clearer understanding of the sample composition.

\begin{table}[htbp]
\centering
\caption{{\centering Distribution by Years of Experience}}
\label{tab:res_01}
\begin{tabular}{lcc}
\hline
\textbf{Years of Experience} & \textbf{Total} & \textbf{\%} \\
\hline
10 or more years & 25 & 50\% \\
4 years & 8 & 16\% \\
5 or more years & 6 & 12\% \\
3 years & 4 & 8\% \\
2 years & 4 & 8\% \\
Up to 1 year & 2 & 4\% \\
1 year & 1 & 2\% \\
\hline
\end{tabular}
\end{table}

\begin{table}[htbp]
\centering
\caption{Area of Expertise of Participants}
\label{tab:res_02}
\begin{tabular}{lcc}
\hline
\textbf{Area of Expertise} & \textbf{Total} & \textbf{\%} \\
\hline
Technology (Infrastructure) & 11 & 22\% \\
Information Security (Defensive) & 11 & 22\% \\
Technology (Software) & 9 & 18\% \\
Information Security (Risk Management) & 5 & 10\% \\
Technology (Data) & 4 & 8\% \\
Others & 3 & 6\% \\
Information Security (Governance) & 3 & 6\% \\
Technology (Governance) & 2 & 4\% \\
Information Security (Offensive) & 2 & 4\% \\
\hline
\end{tabular}
\end{table}

\begin{table}[htbp]
\centering
\caption{Professional Profile of Participants}
\label{tab:res_03}
\begin{tabular}{lcc}
\hline
\textbf{Professional Profile} & \textbf{Total} & \textbf{\%} \\
\hline
Technical & 22 & 44\% \\
Both (Technical and Managerial) & 17 & 34\% \\
Managerial & 11 & 22\% \\
\hline
\end{tabular}
\end{table}

\begin{table}[htbp]
\centering
\caption{Professional Level of Participants}
\label{tab:res_04}
\begin{tabular}{lcc}
\hline
\textbf{Professional Level} & \textbf{Total} & \textbf{\%} \\
\hline
Senior & 19 & 38\% \\
Expert & 13 & 26\% \\
Mid-level & 10 & 20\% \\
Others & 4 & 8\% \\
Intern & 3 & 6\% \\
Junior & 1 & 2\% \\
\hline
\end{tabular}
\end{table}

We grouped participants into four categories, each defined by specific attributes intended to help us understand their risk appetite across the 18 survey scenarios. The categories are as follows:

\begin{enumerate}
    \item Years of experience: Years of experience of these professionals in the area of knowledge of Technology or Information Security;
    \item Area of expertise: Which sub-area of knowledge of Technology or Information Security are they currently working in; 
    \item Professional profile: Professionals are more focused on the technical, managerial, or a mix of both; 
    \item Professional level: The position they occupy, based on the skills, knowledge, or expertise they currently have, which gives them authority and decision-making capacity, described as positions.
\end{enumerate}

The survey, which centers on assigning risk levels to cybersecurity situations, draws on the expertise of 50 professionals, more than half of whom have over 10 years of experience or hold Senior or Expert positions. Additionally, the diverse backgrounds of the participants—spanning Information Security, Software, Data, Infrastructure, and Governance—and their varied roles, whether technical, managerial, or a blend of both, contribute to a comprehensive understanding of how different backgrounds approach risk evaluation. This blend of seasoned expertise and multidisciplinary insights provides a robust foundation for learning and enhancing the assessment of cybersecurity risks. 

\subsection{RQ 1: To what extent can Large Language Models be trusted to conduct cybersecurity risk assessments without the involvement of human experts?} \label{subsection_rq1}

In this study, participants and the five models explored here were assigned a risk rating (0-10) for 18 scenarios based on each of the Center for Internet Security (CIS) controls. By comparing average scores and examining how the ratings of these five models correlate with those of human experts, we gain insight into the overall risk appetite between humans and the LLM models. We then highlight systematic differences—such as the five models' tendency to score lower than human experts—and examine whether the models' relative risk ratings match the patterns observed among human respondents. 

This RQ1 aims to evaluate whether a large language model, such as the five models explored here (ChatGPT 5, DeepSeek V3.1, Llama 4 Scout, Claude Opus 4.1, and Gemini 2.5 PRO), can be relied upon to assess cybersecurity risks without human involvement. To do so, we compare its performance with all surveyed human experts across a set of standardized scenarios. 

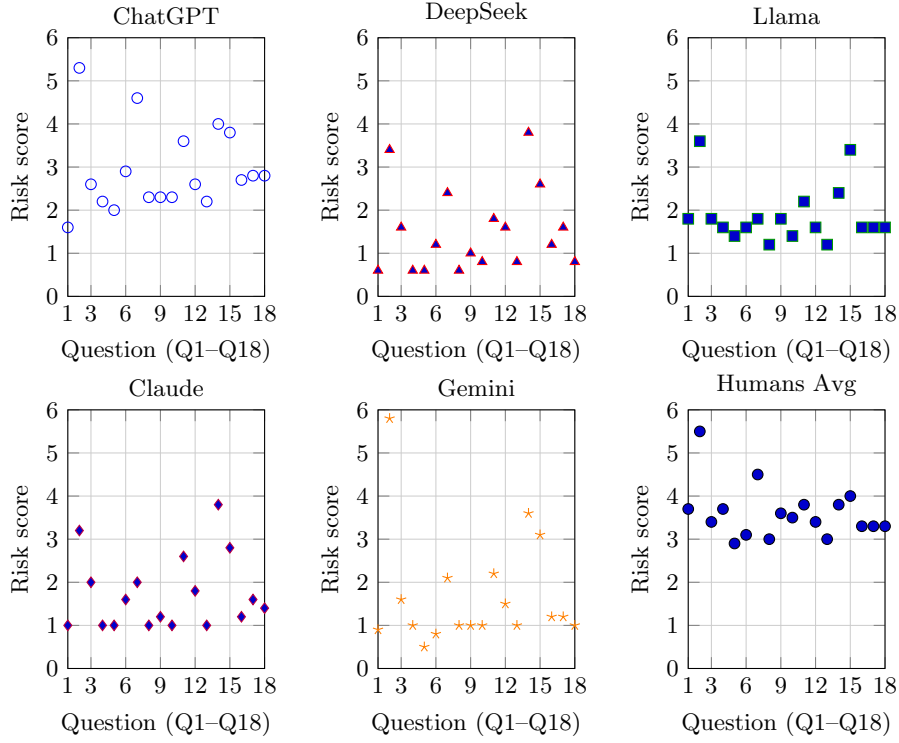
\begin{figure}[!htbp]
\centering
\begin{tikzpicture}
\begin{groupplot}[
  group style={
    group size=3 by 2, % 3 colunas x 2 linhas
    horizontal sep=1.5cm,
    vertical sep=1.5cm
  },
  width=0.32\linewidth,
  height=5cm,
  xlabel={Question (Q1--Q18)},
  ylabel={Risk score},
  xmin=1, xmax=18,
  ymin=0, ymax=6,
  xtick={1,3,6,9,12,15,18},
  ytick={0,1,2,3,4,5,6},
  tick label style={font=\scriptsize},
  title style={font=\small, yshift=-1ex}
]

% GPT
\nextgroupplot[title=ChatGPT]
\addplot+[only marks, mark=o, color=blue] coordinates {
(1,1.60) 
(2,5.30) 
(3,2.60) 
(4,2.20) 
(5,2.00) 
(6,2.90) 
(7,4.60) 
(8,2.30) 
(9,2.30) 
(10,2.30) 
(11,3.60) 
(12,2.60) 
(13,2.20) 
(14,4.00) 
(15,3.80) 
(16,2.70) 
(17,2.80) 
(18,2.80)
};

% DeepSeek
\nextgroupplot[title=DeepSeek]
\addplot+[only marks, mark=triangle*, color=red] coordinates {
(1,0.60)
(2,3.40)
(3,1.60)
(4,0.60)
(5,0.60)
(6,1.20)
(7,2.40)
(8,0.60)
(9,1.00)
(10,0.80)
(11,1.80)
(12,1.60)
(13,0.80)
(14,3.80)
(15,2.60)
(16,1.20)
(17,1.60)
(18,0.80)
};

% Llama
\nextgroupplot[title=Llama]
\addplot+[only marks, mark=square*, color=green!60!black] coordinates {
(1,1.80)
(2,3.60)
(3,1.80)
(4,1.60)
(5,1.40)
(6,1.60)
(7,1.80)
(8,1.20)
(9,1.80)
(10,1.40)
(11,2.20)
(12,1.60)
(13,1.20)
(14,2.40)
(15,3.40)
(16,1.60)
(17,1.60)
(18,1.60)
};

% Claude
\nextgroupplot[title=Claude]
\addplot+[only marks, mark=diamond*, color=purple] coordinates {
(1,1.00)
(2,3.20)
(3,2.00)
(4,1.00)
(5,1.00)
(6,1.60)
(7,2.00)
(8,1.00)
(9,1.20)
(10,1.00)
(11,2.60)
(12,1.80)
(13,1.00)
(14,3.80)
(15,2.80)
(16,1.20)
(17,1.60)
(18,1.40)
};

% Gemini
\nextgroupplot[title=Gemini]
\addplot+[only marks, mark=star, color=orange] coordinates {
(1,0.90)
(2,5.80)
(3,1.60)
(4,1.00)
(5,0.50)
(6,0.80)
(7,2.10)
(8,1.00)
(9,1.00)
(10,1.00)
(11,2.20)
(12,1.50)
(13,1.00)
(14,3.60)
(15,3.10)
(16,1.20)
(17,1.20)
(18,1.00)
};

% Humans
\nextgroupplot[title=Humans Avg]
\addplot+[only marks, mark=*, color=black] coordinates {
(1,3.70)
(2,5.50)
(3,3.40)
(4,3.70)
(5,2.90)
(6,3.10)
(7,4.50)
(8,3.00)
(9,3.60)
(10,3.50)
(11,3.80)
(12,3.40)
(13,3.00)
(14,3.80)
(15,4.00)
(16,3.30)
(17,3.30)
(18,3.30)
};

\end{groupplot}
\end{tikzpicture}
\caption{Scatter plots of risk scores assigned to each question (Q1--Q18) by LLMs and Human Experts.}
\label{fig:scatter-llms-humans}
\end{figure}

First, we analyzed the data by comparing the risk assigned by the five models with that assigned by all other human experts. For each scenario question, we calculated the average risk assigned by the models, the average risk assigned by the human experts, and the standard deviation of the responses. Table \ref{tab:ress_06} shows the average risk assigned to each question by the five models (ChatGPT 5, DeepSeek V3.1, Llama 4 Scout, Claude Opus 4.1, and Gemini 2.5 PRO) vs all human experts surveyed. Figure \ref{fig:scatter-llms-humans} presents the scatter plots of each question per model. 

\begin{table}[!h]
\centering
\caption{Average risk ratings assigned by LLM models and human experts for each scenario question}
\label{tab:ress_06}
\begin{tabular}{lccccc|cc}
\toprule
\textbf{Q} & \textbf{ChatGPT} & \textbf{DeepSeek} & \textbf{Llama} & \textbf{Claude} & \textbf{Gemini} & \textbf{Humans Avg} & \textbf{SD} \\
\midrule
Q1	& 1.60 & 0.60 & 1.80 & 1.00 & 0.90 & 3.70 & 2.45 \\
Q2	& 5.30 & 3.40 & 3.60 & 3.20 & 5.80 & 5.50 & 2.34 \\
Q3	& 2.60 & 1.60 & 1.80 & 2.00 & 1.60 & 3.40 & 2.40 \\
Q4	& 2.20 & 0.60 & 1.60 & 1.00 & 1.00 & 3.70 & 2.66 \\
Q5	& 2.00 & 0.60 & 1.40 & 1.00 & 0.50 & 2.90 & 2.33 \\
Q6	& 2.90 & 1.20 & 1.60 & 1.60 & 0.80 & 3.10 & 2.25 \\
Q7	& 4.60 & 2.40 & 1.80 & 2.00 & 2.10 & 4.50 & 2.74 \\
Q8	& 2.30 & 0.60 & 1.20 & 1.00 & 1.00 & 3.00 & 2.12 \\
Q9	& 2.30 & 1.00 & 1.80 & 1.20 & 1.00 & 3.60 & 2.59 \\
Q10	& 2.30 & 0.80 & 1.40 & 1.00 & 1.00 & 3.50 & 2.70 \\
Q11	& 3.60 & 1.80 & 2.20 & 2.60 & 2.20 & 3.80 & 2.44 \\
Q12	& 2.60 & 1.60 & 1.60 & 1.80 & 1.50 & 3.40 & 2.40 \\
Q13	& 2.20 & 0.80 & 1.20 & 1.00 & 1.00 & 3.00 & 2.06 \\
Q14	& 4.00 & 3.80 & 2.40 & 3.80 & 3.60 & 3.80 & 2.61 \\
Q15	& 3.80 & 2.60 & 3.40 & 2.80 & 3.10 & 4.00 & 2.30 \\
Q16	& 2.70 & 1.20 & 1.60 & 1.20 & 1.20 & 3.30 & 2.53 \\
Q17	& 2.80 & 1.60 & 1.60 & 1.60 & 1.20 & 3.30 & 2.44 \\
Q18	& 2.80 & 0.80 & 1.60 & 1.40 & 1.00 & 3.30 & 2.48 \\
\midrule
\textbf{Total} & \textbf{2.92} & \textbf{1.50} & \textbf{1.87} & \textbf{1.73} & \textbf{1.69} & \textbf{3.60} & \textbf{2.44} \\
\bottomrule
\end{tabular}
\end{table}

Our results indicate that the overall average risk assigned by the human experts is higher than that designated by the five models for all questions. The average risk assigned by the models for the 18 questions is \textbf{1.94}, while the average risk assigned by the human experts to the same 18 questions is \textbf{3.60}. Human experts assign a higher risk with a standard deviation of \textbf{2.44}. In the overall average across the five models and human experts, models estimate a risk that is \textbf{1.66 lower} than human experts.

%%%%%%%%%%%%%%%%%%%%%%%%%%%%%%%%%%%%%%%%%%%%%%%%%%%%%%%%%%%%%
% Fig 3 - Gráfico de dispersão RQ1 All models vs All Humans
%%%%%%%%%%%%%%%%%%%%%%%%%%%%%%%%%%%%%%%%%%%%%%%%%%%%%%%%%%%%%

Second, we wanted to know the correlation coefficient between the five models and human experts. In practical terms, is there a direct correlation between the risk ratings assigned by models and the human experts? That is, do the risk ratings move in the same direction? To understand this better, we use Pearson's correlation coefficient to measure the strength of the linear relationship between two variables. If there is a strong linear relationship, the correlation coefficient is close to 1 or $-1$; 0 indicates no linear relationship.  

The mean Pearson correlation coefficient for the five models of \textbf{0.81} shows that although models are lower on average, it still moves in the same direction as human ratings to a moderate to moderately strong positive correlation degree (if humans score a particular scenario higher, the five models tends to do so too, but not as high in absolute terms), as shown in Table \ref{tab:pearson-corr-geral}. 

%%%%%%%%%%%%%%%%%%%%%%%%%%%%%%%%%%%%%%%%%%%%%%%%%%%%%%%%%%%%%
% Tabela 7 Sumarios de coeficientes Pearson All Models vs All Humans
%%%%%%%%%%%%%%%%%%%%%%%%%%%%%%%%%%%%%%%%%%%%%%%%%%%%%%%%%%%%%
\begin{table}[!htbp]
\centering
\caption{Pearson correlation coefficients between human experts and LLM models}
\label{tab:pearson-corr-geral}
\begin{tabular}{lcccccc}
\toprule
& \textbf{LLMs Average} & \textbf{ChatGPT} & \textbf{DeepSeek} & \textbf{Llama} & \textbf{Claude} & \textbf{Gemini} \\
\midrule
Humans & 0.814 & 0.817 & 0.706 & 0.801 & 0.630 & 0.858 \\
\bottomrule
\end{tabular}
\end{table}

Third, we conducted statistical tests to evaluate the significance of the differences using a paired-samples t-test for the five models and human experts \cite{rs02}. The paired-samples t-test, also known as the dependent-samples t-test, is a statistical test for determining whether the mean difference between two observational data sets equals zero. Each object or entity is measured twice in this test, resulting in two sets of observations. The paired-samples t-test employs two contradictory research hypotheses, the null and alternative hypotheses, as do most statistical procedures. The null hypothesis states that there is no difference in the means of the two paired datasets. According to this perspective, all visible distinctions result from random variation. It is possible that the actual mean difference between the two samples is not equal to zero, contrary to the alternative hypothesis.

\begin{table}[!htbp]
\centering
\caption{Paired t-test results comparing human experts with LLM models}
\label{tab:ttest}
\begin{tabular}{lcc}
\toprule
\textbf{Model} & \textbf{t Stat} & \textbf{p-value (two-tail)} \\
\midrule
LLMs Average & 12.94 & 3.14$\times 10^{-10}$ \\
GPT          & 4.91 & 1.32$\times 10^{-4}$ \\
DeepSeek     & 12.71 & 4.16$\times 10^{-10}$ \\
Llama        & 17.59 & 2.41$\times 10^{-12}$ \\
Claude       & 11.69 & 1.49$\times 10^{-9}$ \\
Gemini       & 9.59  & 2.83$\times 10^{-8}$ \\
\bottomrule
\end{tabular}
\end{table}

Since all p-values obtained from the paired t-tests are extremely small \textbf{(all $p < 0.001$)}, we can confidently reject the null hypothesis in every case. This result indicates that, despite the positive correlations between the five models' predictions and human ratings, there are statistically significant differences in the average risk ratings assigned by each model compared to those of human experts. In other words, while the models may follow similar trends, their absolute risk assessments systematically diverge from those provided by humans.

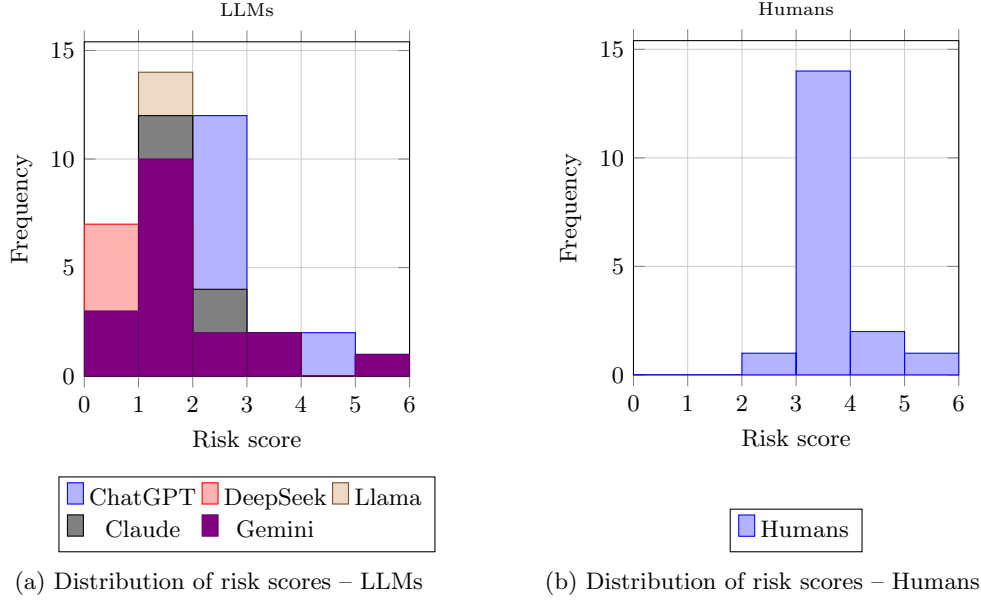
\begin{figure}[!htbp]
\centering
%%%%%%%%%%%%%%%%%%%%%%%%%%%%%%%%%%%%%%%%%%%%%%%%%%%%%%%%%%%%%
% Subfigure Historam LLM
%%%%%%%%%%%%%%%%%%%%%%%%%%%%%%%%%%%%%%%%%%%%%%%%%%%%%%%%%%%%%
\begin{subfigure}[t]{0.45\textwidth}
\centering
\begin{tikzpicture}
\begin{axis}[
  ybar,
  bar width=3pt,
  width=\linewidth,
  height=6cm,
  xlabel={Risk score},
  ylabel={Frequency},
  ymin=0,
  xmin=0, xmax=6,
  xtick={0,1,2,3,4,5,6},
  title={LLMs},
  legend style={
    at={(0.5,-0.3)},
    anchor=north,
    legend columns=3,
    font=\scriptsize}
]
% GPT
\addplot+[
  hist={bins=6, data min=0, data max=6}
] table [y index=0, row sep=\\] {data\\
1.60\\ 5.30\\ 2.60\\ 2.20\\ 2.00\\ 2.90\\ 4.60\\ 2.30\\ 2.30\\ 2.30\\ 3.60\\ 2.60\\ 2.20\\ 4.00\\ 3.80\\ 2.70\\ 2.80\\ 2.80\\};
\addlegendentry{ChatGPT}

% DeepSeek
\addplot+[
  hist={bins=6, data min=0, data max=6}
] table [y index=0, row sep=\\] {data\\
0.60\\ 3.40\\ 1.60\\ 0.60\\ 0.60\\ 1.20\\ 2.40\\ 0.60\\ 1.00\\ 0.80\\ 1.80\\ 1.60\\ 0.80\\ 3.80\\ 2.60\\ 1.20\\ 1.60\\ 0.80\\};
\addlegendentry{DeepSeek}

% Llama
\addplot+[
  hist={bins=6, data min=0, data max=6}
] table [y index=0, row sep=\\] {data\\
1.80\\ 3.60\\ 1.80\\ 1.60\\ 1.40\\ 1.60\\ 1.80\\ 1.20\\ 1.80\\ 1.40\\ 2.20\\ 1.60\\ 1.20\\ 2.40\\ 3.40\\ 1.60\\ 1.60\\ 1.60\\};
\addlegendentry{Llama}

% Claude
\addplot+[
  hist={bins=6, data min=0, data max=6}
] table [y index=0, row sep=\\] {data\\
1.00\\ 3.20\\ 2.00\\ 1.00\\ 1.00\\ 1.60\\ 2.00\\ 1.00\\ 1.20\\ 1.00\\ 2.60\\ 1.80\\ 1.00\\ 3.80\\ 2.80\\ 1.20\\ 1.60\\ 1.40\\};
\addlegendentry{Claude}

% Gemini
\addplot+[
  hist={bins=6, data min=0, data max=6}
] table [y index=0, row sep=\\] {data\\
0.90\\ 5.80\\ 1.60\\ 1.00\\ 0.50\\ 0.80\\ 2.10\\ 1.00\\ 1.00\\ 1.00\\ 2.20\\ 1.50\\ 1.00\\ 3.60\\ 3.10\\ 1.20\\ 1.20\\ 1.00\\ };
\addlegendentry{Gemini}

\end{axis}
\end{tikzpicture}
\caption{Distribution of risk scores – LLMs}
\end{subfigure}%
\hfill
%%%%%%%%%%%%%%%%%%%%%%%%%%%%%%%%%%%%%%%%%%%%%%%%%%%%%%%%%%%%%
% Subfigure Historam Humans
%%%%%%%%%%%%%%%%%%%%%%%%%%%%%%%%%%%%%%%%%%%%%%%%%%%%%%%%%%%%%
\begin{subfigure}[t]{0.45\textwidth}
\centering
\begin{tikzpicture}
\begin{axis}[
  ybar,
  bar width=12pt,
  width=\linewidth,
  height=6cm,
  xlabel={Risk score},
  ylabel={Frequency},
  ymin=0,
  xmin=0, xmax=6,
  xtick={0,1,2,3,4,5,6},
  title={Humans},
  legend style={
    at={(0.5,-0.4)},
    anchor=north,
    legend columns=3,
    font=\scriptsize}
]
\addplot+[
  hist={bins=6, data min=0, data max=6}
] table [y index=0, row sep=\\] {data\\
3.7\\ 5.5\\ 3.4\\ 3.7\\ 2.9\\ 3.1\\ 4.5\\ 3.0\\ 3.6\\ 3.5\\ 3.8\\ 3.4\\ 3.0\\ 3.8\\ 4.0\\ 3.3\\ 3.3\\ 3.3\\};
\addlegendentry{Humans}
\end{axis}
\end{tikzpicture}
\caption{Distribution of risk scores – Humans}
\end{subfigure}

\caption{Comparison of distributions of risk scores assigned by LLMs and Human Experts (Q1–Q18).}
\label{fig:hist-llm-vs-human}
\end{figure}

Finally, the histograms show the distribution of risk ratings assigned by the five models versus those given by human participants (Figure \ref{fig:hist-llm-vs-human}). While human experts predominantly rated the scenarios as medium-to-high risk (scores 3-5), LLMs concentrated their assessments in the lower range (1–3), with only occasional higher values. This indicates that, although LLMs follow similar trends to humans, they consistently underestimate the overall risk level, reinforcing the t-test results.
%%%%%%%%%%%%%%%%%%%%%%%%%%%%%%%%%%%%%%%%%%%%%%%%%%%%%%%%%%%%%
% Fig 4 Histrogram of LLM
%%%%%%%%%%%%%%%%%%%%%%%%%%%%%%%%%%%%%%%%%%%%%%%%%%%%%%%%%%%%%

The findings discussed above support the conclusion that the five models systematically underestimate cybersecurity risk compared to the perception of all human participants. In addition, the standard deviation of \textbf{2.44} across human participants indicates a significant distribution of opinions. While some participants may rate a scenario as very low risk, others may rate it as higher risk. This variation suggests that the group of human participants does not have a rigid consensus on the scenarios, reflecting a wide range of perspectives or interpretations when judging risk, which will be more detailed in the following sections. 
%%%%%%%%%%%%%%%%%%%%%%%%%%%%%%%%%%%%%%%%%%%%%%%%%%%%%%%%%%%%%
% SUB-Seção RQ2 - Tópico 5.3
%%%%%%%%%%%%%%%%%%%%%%%%%%%%%%%%%%%%%%%%%%%%%%%%%%%%%%%%%%%%%
\subsection{RQ 2. How does the Large Language Model (LLM) used in this study compare to human cybersecurity professionals in assessing risk across standardized scenarios} \label{subsection_rq2}

This research question aims to investigate the extent to which a large language model can replicate or approximate cybersecurity professionals' judgments when performing structured risk assessments—one of the attributes used in this study to group human participants into categories. Professional Level is one category comprising two large groups of people, with \textbf{Senior and Expert participants accounting for 64\%} of the sample and \textbf{all other levels accounting for the remaining 36\%}. Given the diversity of expertise within the cybersecurity workforce, this question is further divided into two sub-questions that reflect different levels of professional experience, RQ2.1 (\ref{rq21}) and RQ2.2 (\ref{rq22}):

%%%%%%%%%%%%%%%%%%%%%%%%%%%%%%%%%%%%%%%%%%%%%%%%%%%%%%%%%%%%%
% SUB-Seção RQ2.1 - Tópico 5.3.1
%%%%%%%%%%%%%%%%%%%%%%%%%%%%%%%%%%%%%%%%%%%%%%%%%%%%%%%%%%%%%
\subsubsection{How does the five models performance compare to that of senior and specialist professionals, who are expected to have deeper technical knowledge and more refined risk perception?} \label{rq21}

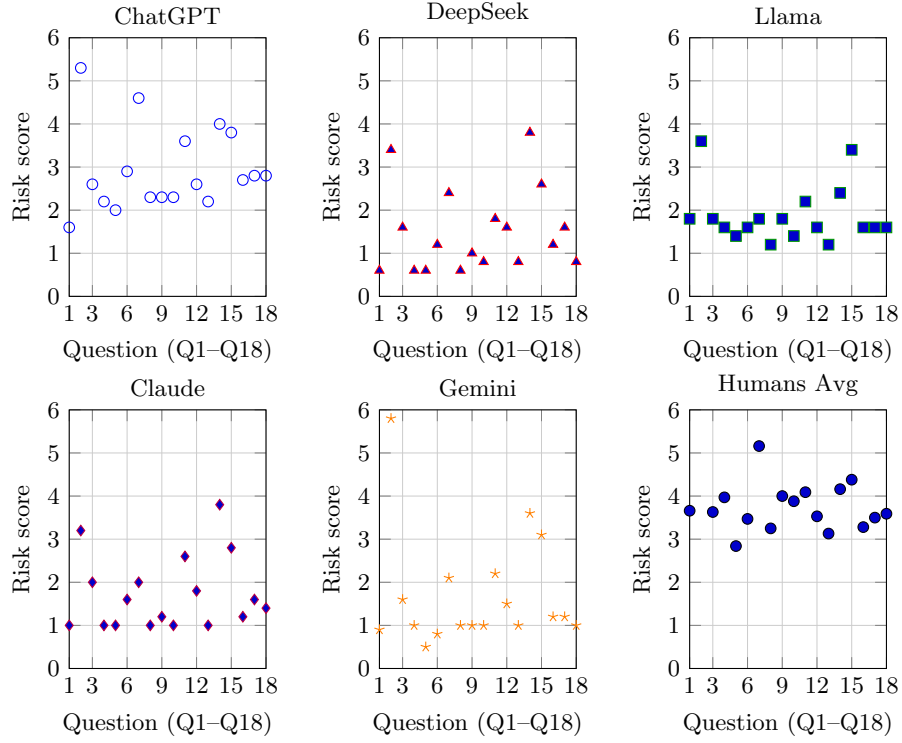
\begin{figure}[!h]
\centering
\begin{tikzpicture}
\begin{groupplot}[
  group style={
    group size=3 by 2, % 3 colunas x 2 linhas
    horizontal sep=1.5cm,
    vertical sep=1.5cm
  },
  width=0.32\linewidth,
  height=5cm,
  xlabel={Question (Q1--Q18)},
  ylabel={Risk score},
  xmin=1, xmax=18,
  ymin=0, ymax=6,
  xtick={1,3,6,9,12,15,18},
  ytick={0,1,2,3,4,5,6},
  tick label style={font=\scriptsize},
  title style={font=\small, yshift=-1ex}
]

% GPT
\nextgroupplot[title=ChatGPT]
\addplot+[only marks, mark=o, color=blue] coordinates {
(1,1.60)
(2,5.30)
(3,2.60)
(4,2.20)
(5,2.00)
(6,2.90)
(7,4.60)
(8,2.30)
(9,2.30)
(10,2.30)
(11,3.60)
(12,2.60)
(13,2.20)
(14,4.00)
(15,3.80)
(16,2.70)
(17,2.80)
(18,2.80)
};

% DeepSeek
\nextgroupplot[title=DeepSeek]
\addplot+[only marks, mark=triangle*, color=red] coordinates {
(1,0.60)
(2,3.40)
(3,1.60)
(4,0.60)
(5,0.60)
(6,1.20)
(7,2.40)
(8,0.60)
(9,1.00)
(10,0.80)
(11,1.80)
(12,1.60)
(13,0.80)
(14,3.80)
(15,2.60)
(16,1.20)
(17,1.60)
(18,0.80)
};

% Llama
\nextgroupplot[title=Llama]
\addplot+[only marks, mark=square*, color=green!60!black] coordinates {
(1,1.80)
(2,3.60)
(3,1.80)
(4,1.60)
(5,1.40)
(6,1.60)
(7,1.80)
(8,1.20)
(9,1.80)
(10,1.40)
(11,2.20)
(12,1.60)
(13,1.20)
(14,2.40)
(15,3.40)
(16,1.60)
(17,1.60)
(18,1.60)
};

% Claude
\nextgroupplot[title=Claude]
\addplot+[only marks, mark=diamond*, color=purple] coordinates {
(1,1.00)
(2,3.20)
(3,2.00)
(4,1.00)
(5,1.00)
(6,1.60)
(7,2.00)
(8,1.00)
(9,1.20)
(10,1.00)
(11,2.60)
(12,1.80)
(13,1.00)
(14,3.80)
(15,2.80)
(16,1.20)
(17,1.60)
(18,1.40)
};

% Gemini
\nextgroupplot[title=Gemini]
\addplot+[only marks, mark=star, color=orange] coordinates {
(1,0.90)
(2,5.80)
(3,1.60)
(4,1.00)
(5,0.50)
(6,0.80)
(7,2.10)
(8,1.00)
(9,1.00)
(10,1.00)
(11,2.20)
(12,1.50)
(13,1.00)
(14,3.60)
(15,3.10)
(16,1.20)
(17,1.20)
(18,1.00)
};

% Humans
\nextgroupplot[title=Humans Avg]
\addplot+[only marks, mark=*, color=black] coordinates {
(1,3.66)
(2,6.22)
(3,3.63)
(4,3.97)
(5,2.84)
(6,3.47)
(7,5.16)
(8,3.25)
(9,4.00)
(10,3.88)
(11,4.09)
(12,3.53)
(13,3.13)
(14,4.16)
(15,4.38)
(16,3.28)
(17,3.50)
(18,3.59)
};

\end{groupplot}
\end{tikzpicture}
\caption{Scatter plots of risk scores assigned to each question (Q1--Q18) by  LLMs with Senior \& Specialist Human Experts.}
\label{fig:scatter-llms-humans-sen-spec}
\end{figure}

First, we analyzed the data by comparing the risk assigned by the five models with that assigned by human experts. This time, only the group of participants with Senior and Specialist professional levels. For each scenario question, we calculated the average risk assigned by the five models, the average risk assigned by the Senior and Specialist Professionals, and the standard deviation of their responses. The results in Table \ref{tab:avg-risk-senior-specialist} show each participant's average risk assigned to each question: Five models vs. All Senior and Specialist Professionals. Figure \ref{fig:scatter-llms-humans-sen-spec} presents the scatter plots for each model. 

%%%%%%%%%%%%%%%%%%%%%%%%%%%%%%%%%%%%%%%%%%%%%%%%%%%%%%%%%%%%%
% Tabela 9 Average Risk by Todos LLM vs Senior e Specs
%%%%%%%%%%%%%%%%%%%%%%%%%%%%%%%%%%%%%%%%%%%%%%%%%%%%%%%%%%%%%
\begin{table}[!h]
\centering
\caption{Average risk ratings assigned by LLM Models vs Senior and Specialist Human Experts for each scenario question.}
\label{tab:avg-risk-senior-specialist}
\setlength{\tabcolsep}{5pt}
\begin{tabular*}{\linewidth}{@{\extracolsep{\fill}} l c c c c c !{\vrule width 0.3pt} c c }
\toprule
\textbf{Q} & \textbf{ChatGPT} & \textbf{DeepSeek} & \textbf{Llama} & \textbf{Claude} & \textbf{Gemini} & \textbf{Human Mean} & \textbf{SD} \\
\midrule
Q1	& 1.60 & 0.60 & 1.80 & 1.00 & 0.90 & 3.66 & 1.94 \\
Q2	& 5.30 & 3.40 & 3.60 & 3.20 & 5.80 & 6.22 & 2.05 \\
Q3	& 2.60 & 1.60 & 1.80 & 2.00 & 1.60 & 3.63 & 2.44 \\
Q4	& 2.20 & 0.60 & 1.60 & 1.00 & 1.00 & 3.97 & 2.53 \\
Q5	& 2.00 & 0.60 & 1.40 & 1.00 & 0.50 & 2.84 & 2.08 \\
Q6	& 2.90 & 1.20 & 1.60 & 1.60 & 0.80 & 3.47 & 2.35 \\
Q7	& 4.60 & 2.40 & 1.80 & 2.00 & 2.10 & 5.16 & 2.74 \\
Q8	& 2.30 & 0.60 & 1.20 & 1.00 & 1.00 & 3.25 & 2.32 \\
Q9	& 2.30 & 1.00 & 1.80 & 1.20 & 1.00 & 4.00 & 2.61 \\
Q10	& 2.30 & 0.80 & 1.40 & 1.00 & 1.00 & 3.88 & 2.87 \\
Q11	& 3.60 & 1.80 & 2.20 & 2.60 & 2.20 & 4.09 & 2.34 \\
Q12	& 2.60 & 1.60 & 1.60 & 1.80 & 1.50 & 3.53 & 2.35 \\
Q13	& 2.20 & 0.80 & 1.20 & 1.00 & 1.00 & 3.13 & 2.04 \\
Q14	& 4.00 & 3.80 & 2.40 & 3.80 & 3.60 & 4.16 & 2.80 \\
Q15	& 3.80 & 2.60 & 3.40 & 2.80 & 3.10 & 4.38 & 2.23 \\
Q16	& 2.70 & 1.20 & 1.60 & 1.20 & 1.20 & 3.28 & 2.46 \\
Q17	& 2.80 & 1.60 & 1.60 & 1.60 & 1.20 & 3.50 & 2.48 \\
Q18	& 2.80 & 0.80 & 1.60 & 1.40 & 1.00 & 3.59 & 2.61 \\
\midrule
\textbf{Total} & \textbf{2.92} & \textbf{1.50} & \textbf{1.87} & \textbf{1.73} & \textbf{1.69} & \textbf{3.87} & \textbf{2.41} \\
\bottomrule
\end{tabular*}
\end{table}

Our results show that the overall average risk attributed by this group of Senior and Specialist Human Experts is higher than that attributed by the five models across all questions, and higher than that attributed by all participants. In addition to the average risk of \textbf{3.87}, a smaller standard deviation of \textbf{2.41} is observed compared to the group containing all participants, indicating better consensus within this group. 

%%%%%%%%%%%%%%%%%%%%%%%%%%%%%%%%%%%%%%%%%%%%%%%%%%%%%%%%%%%%%
% Fig 6 - Gráfico de dispersão RQ1 All models vs SR e SPEC
%%%%%%%%%%%%%%%%%%%%%%%%%%%%%%%%%%%%%%%%%%%%%%%%%%%%%%%%%%%%%

Second, we aimed to determine the correlation between the risk ratings assigned by the five models (ChatGPT 5, DeepSeek V3.1, Llama 4 Scout, Claude Opus 4.1, and Gemini 2.5 PRO) and those provided by human experts. Specifically, we sought to assess whether there is a direct relationship, i.e., whether the risk ratings tend to move in the same direction. To examine this, we employed Pearson's correlation coefficient to measure the strength of the linear relationship between the two sets of ratings. A coefficient close to 1 or $-1$ indicates a strong linear relationship, while a value near 0 suggests no linear correlation.

The mean Pearson correlation of \textbf{0.819} shows that although the five models is lower on average, it still moves in the same direction as human ratings to a moderate to moderately strong positive correlation degree (if humans score a particular scenario higher. The five models tends to do so too—but not as high in absolute terms), as shown in Table \ref{tab:pearson-corr-profes-level}.

%%%%%%%%%%%%%%%%%%%%%%%%%%%%%%%%%%%%%%%%%%%%%%%%%%%%%%%%%%%%%
% Tabela 10 Sumarios de coeficientes Pearson All Models vs SR e SPEC
%%%%%%%%%%%%%%%%%%%%%%%%%%%%%%%%%%%%%%%%%%%%%%%%%%%%%%%%%%%%%
\begin{table}[!h]
\centering
\caption{Pearson correlation coefficients between Seniors and Specialist humans experts and LLM models}
\label{tab:pearson-corr-profes-level}
\begin{tabular}{lcccccc}
\toprule
& \textbf{LLMs Average} & \textbf{ChatGPT} & \textbf{DeepSeek} & \textbf{Llama} & \textbf{Claude} & \textbf{Gemini} \\
\midrule
Humans & 0.819 & 0.853 & 0.716 & 0.775 & 0.639 & 0.847 \\
\bottomrule
\end{tabular}
\end{table}

Third, we conducted statistical tests to evaluate the significance of the differences using paired-samples t-tests for the five models and human experts. It shows again a very small p-value of \textbf{(all $p < 0.001$)}, so we can confidently reject the null hypothesis and assume the alternative hypothesis: there is a significant difference in the average risk rating assigned between the two groups.

%%%%%%%%%%%%%%%%%%%%%%%%%%%%%%%%%%%%%%%%%%%%%%%%%%%%%%%%%%%%%
% Tabela 11 Testes pareados, T-test SR e SPEC
%%%%%%%%%%%%%%%%%%%%%%%%%%%%%%%%%%%%%%%%%%%%%%%%%%%%%%%%%%%%%
\begin{table}[!h]
\centering
\caption{Paired t-test results comparing LLM models with Senior \& Specialist human experts (Q1--Q18).}
\label{tab:ttest-llm-humansproflevel}
\begin{tabular}{lcc}
\toprule
\textbf{Model} & \textbf{t Stat} & \textbf{p-value (two-tail)} \\
\midrule
LLMs Average & 15.74 & 1.44$\times 10^{-11}$ \\
GPT      & 7.96 & 3.90$\times 10^{-7}$ \\
DeepSeek & 14.65 & 4.50$\times 10^{-11}$ \\ 
Llama    & 17.03 & 4.06$\times 10^{-12}$ \\
Claude   & 12.97 & 3.01 $\times 10^{-10}$ \\
Gemini   & 12.03 & 9.64$\times 10^{-10}$ \\
\bottomrule
\end{tabular}
\end{table}

The paired t-tests (Table~\ref{tab:ttest-llm-humansproflevel}) show that all LLMs differ significantly from Senior \& Specialist Human Experts \textbf{($p < 0.001$)}. While the models broadly follow similar trends, their mean ratings are consistently lower, confirming that LLMs systematically underestimate risk compared to expert assessments.

%%%%%%%%%%%%%%%%%%%%%%%%%%%%%%%%%%%%%%%%%%%%%%%%%%%%%%%%
%INICIO DA CONCLUSAO DA RQ21 SENIOR E SPEC
%%%%%%%%%%%%%%%%%%%%%%%%%%%%%%%%%%%%%%%%%%%%%%%%%%%%%%%%

\begin{figure}[!htbp]
\centering
%%%%%%%%%%%%%%%%%%%%%%%%%%%%%%%%%%%%%%%%%%%%%%%%%%%%%%%%%%%%%
% Subfigure Historam LLM
%%%%%%%%%%%%%%%%%%%%%%%%%%%%%%%%%%%%%%%%%%%%%%%%%%%%%%%%%%%%%
\begin{subfigure}[t]{0.45\textwidth}
\centering
\begin{tikzpicture}
\begin{axis}[
  ybar,
  bar width=3pt,
  width=\linewidth,
  height=6cm,
  xlabel={Risk score},
  ylabel={Frequency},
  ymin=0,
  xmin=0, xmax=6,
  xtick={0,1,2,3,4,5,6},
  title={LLMs},
  legend style={
    at={(0.5,-0.3)},
    anchor=north,
    legend columns=3,
    font=\scriptsize}
]
% GPT
\addplot+[
  hist={bins=6, data min=0, data max=6}
] table [y index=0, row sep=\\] {data\\
1.60\\ 5.30\\ 2.60\\ 2.20\\ 2.00\\ 2.90\\ 4.60\\ 2.30\\ 2.30\\ 2.30\\ 3.60\\ 2.60\\ 2.20\\ 4.00\\ 3.80\\ 2.70\\ 2.80\\ 2.80\\};
\addlegendentry{ChatGPT}

% DeepSeek
\addplot+[
  hist={bins=6, data min=0, data max=6}
] table [y index=0, row sep=\\] {data\\
0.60\\ 3.40\\ 1.60\\ 0.60\\ 0.60\\ 1.20\\ 2.40\\ 0.60\\ 1.00\\ 0.80\\ 1.80\\ 1.60\\ 0.80\\ 3.80\\ 2.60\\ 1.20\\ 1.60\\ 0.80\\};
\addlegendentry{DeepSeek}

% Llama
\addplot+[
  hist={bins=6, data min=0, data max=6}
] table [y index=0, row sep=\\] {data\\
1.80\\ 3.60\\ 1.80\\ 1.60\\ 1.40\\ 1.60\\ 1.80\\ 1.20\\ 1.80\\ 1.40\\ 2.20\\ 1.60\\ 1.20\\ 2.40\\ 3.40\\ 1.60\\ 1.60\\ 1.60\\};
\addlegendentry{Llama}

% Claude
\addplot+[
  hist={bins=6, data min=0, data max=6}
] table [y index=0, row sep=\\] {data\\
1.00\\ 3.20\\ 2.00\\ 1.00\\ 1.00\\ 1.60\\ 2.00\\ 1.00\\ 1.20\\ 1.00\\ 2.60\\ 1.80\\ 1.00\\ 3.80\\ 2.80\\ 1.20\\ 1.60\\ 1.40\\};
\addlegendentry{Claude}

% Gemini
\addplot+[
  hist={bins=6, data min=0, data max=6}
] table [y index=0, row sep=\\] {data\\
0.90\\ 5.80\\ 1.60\\ 1.00\\ 0.50\\ 0.80\\ 2.10\\ 1.00\\ 1.00\\ 1.00\\ 2.20\\ 1.50\\ 1.00\\ 3.60\\ 3.10\\ 1.20\\ 1.20\\ 1.00\\ };
\addlegendentry{Gemini}

\end{axis}
\end{tikzpicture}
\caption{Distribution of risk scores – LLMs}
\end{subfigure}%
\hfill
%%%%%%%%%%%%%%%%%%%%%%%%%%%%%%%%%%%%%%%%%%%%%%%%%%%%%%%%%%%%%
% Subfigure Historam Humans
%%%%%%%%%%%%%%%%%%%%%%%%%%%%%%%%%%%%%%%%%%%%%%%%%%%%%%%%%%%%%
\begin{subfigure}[t]{0.45\textwidth}
\centering
\begin{tikzpicture}
\begin{axis}[
  ybar,
  bar width=12pt,
  width=\linewidth,
  height=6cm,
  xlabel={Risk score},
  ylabel={Frequency},
  ymin=0,
  xmin=0, xmax=6,
  xtick={0,1,2,3,4,5,6},
  title={Humans},
  legend style={
    at={(0.5,-0.4)},
    anchor=north,
    legend columns=3,
    font=\scriptsize}
]
\addplot+[
  hist={bins=6, data min=0, data max=6}
] table [y index=0, row sep=\\] {data\\
3.66\\ 6.22\\ 3.63\\ 3.97\\ 2.84\\ 3.47\\ 5.16\\ 3.25\\ 4.00\\ 3.88\\ 4.09\\ 3.53\\ 3.13\\ 4.16\\ 4.38\\ 3.28\\ 3.50\\ 3.59\\};
\addlegendentry{Humans}
\end{axis}
\end{tikzpicture}
\caption{Distribution of risk scores – Humans}
\end{subfigure}

\caption{Comparison of distributions of risk scores assigned by LLMs and Human Experts (Q1–Q18).}
\label{fig:hist-llm-vs-human-proflevel}
\end{figure}
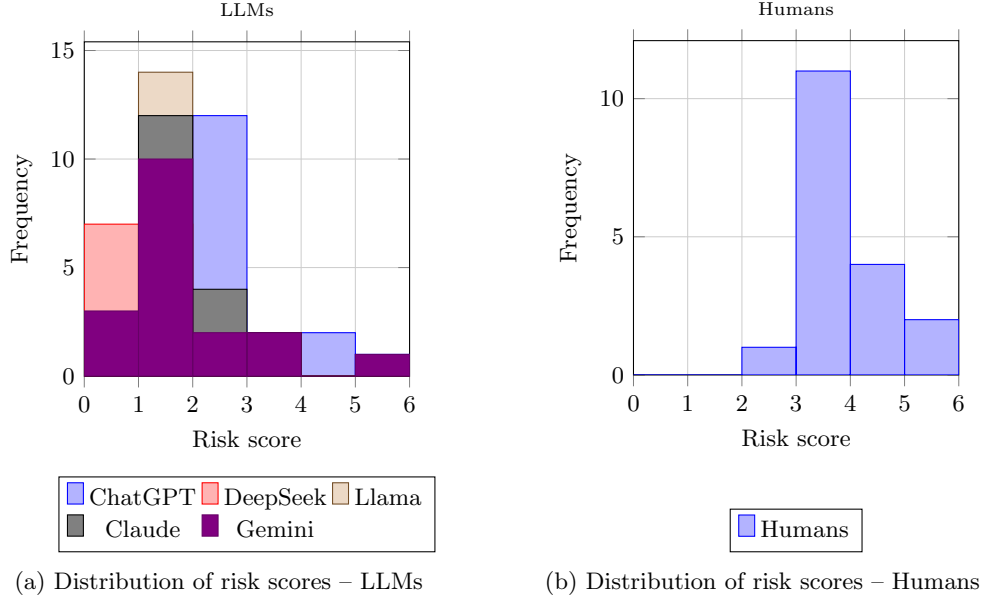

Finally, we analyzed histograms to compare the distribution of risk ratings assigned by the five models with those provided by human participants. The histogram of Senior \& Specialist Human Experts’ ratings (Figure~\ref{fig:hist-llm-vs-human-proflevel}) shows a clear concentration of scores in the medium-to-high risk range (between 3 and 6), indicating a consistent perception of elevated risk across most scenarios and very few ratings below 3. In contrast with the LLM histograms—typically skewed toward lower-risk values—the experts’ distribution shows a higher central tendency (mean $\approx$ 3.87) and reduced variability at extreme values. These differences highlight the tendency of human experts to adopt a more cautious stance when assessing potential risks, reinforcing the statistical gap observed between automated model outputs and professional judgment.

%%%%%%%%%%%%%%%%%%%%%%%%%%%%%%%%%%%%%%%%%%%%%%%%%%%%%%%%%%%%%
% Fig 6 Histrogram of LLM Senior e Spec
%%%%%%%%%%%%%%%%%%%%%%%%%%%%%%%%%%%%%%%%%%%%%%%%%%%%%%%%%%%%%

%%%%%%%%%%%%%%%%%%%%%%%%%%%%%%%%%%%%%%%%%%%%%%%%%%%%%%%%%
% RQ21 Seniors and Specs CONCLUSAO FINAL
%%%%%%%%%%%%%%%%%%%%%%%%%%%%%%%%%%%%%%%%%%%%%%%%%%%%%%%%%
In contrast, the five models' distributions remain heavily concentrated in the lowest bins, with only a handful of ratings above 3. These results highlight how the five models continue to underestimate risk relative to the broader, more varied assessments made by human respondents. 

Our findings support the conclusion that the five models continue to underestimate cybersecurity risk relative to human participants' perceptions systematically. This difference is even more pronounced when compared explicitly with the Senior and Specialist group, which consistently assigned higher risk scores than models and demonstrated better agreement among themselves, as reflected in the lower standard deviation in their responses.

%%%%%%%%%%%%%%%%%%%%%%%%%%%%%%%%%%%%%%%%%%%%%%%%%%%%%%%%%
%%%%%%%%%%%%%%%%%%%%%%%%%%%%%%%%%%%%%%%%%%%%%%%%%%%%%%%%%
%INICIO DA CONCLUSAO DA RQ22 OTHER LEVELS
%%%%%%%%%%%%%%%%%%%%%%%%%%%%%%%%%%%%%%%%%%%%%%%%%%%%%%%%%
%%%%%%%%%%%%%%%%%%%%%%%%%%%%%%%%%%%%%%%%%%%%%%%%%%%%%%%%%
\subsubsection{\textbf{RQ2.2:} How does the LLMs performance compare to other professionals, including junior, mid-level, and non-specialist participants? \label{rq22}
}
First, we analyzed the data by comparing the risk assigned by the five models with that assigned by Human Experts, this time only the group of other professional levels (Non-Senior/Specialist). For each scenario question, we calculated the average risk assigned by the five models, the average risk assigned by the different professional levels, and the standard deviation of their responses. 

\begin{figure}[!htbp]
\centering
\begin{tikzpicture}
\begin{groupplot}[
  group style={
    group size=3 by 2, % 3 colunas x 2 linhas
    horizontal sep=1.5cm,
    vertical sep=1.5cm
  },
  width=0.32\linewidth,
  height=5cm,
  xlabel={Question (Q1--Q18)},
  ylabel={Risk score},
  xmin=1, xmax=18,
  ymin=0, ymax=6,
  xtick={1,3,6,9,12,15,18},
  ytick={0,1,2,3,4,5,6},
  tick label style={font=\scriptsize},
  title style={font=\small, yshift=-1ex}
]

% GPT
\nextgroupplot[title=ChatGPT]
\addplot+[only marks, mark=o, color=blue] coordinates {
(1,1.60)
(2,5.30)
(3,2.60)
(4,2.20)
(5,2.00)
(6,2.90)
(7,4.60)
(8,2.30)
(9,2.30)
(10,2.30)
(11,3.60)
(12,2.60)
(13,2.20)
(14,4.00)
(15,3.80)
(16,2.70)
(17,2.80)
(18,2.80)
};

% DeepSeek
\nextgroupplot[title=DeepSeek]
\addplot+[only marks, mark=triangle*, color=red] coordinates {
(1,0.60)
(2,3.40)
(3,1.60)
(4,0.60)
(5,0.60)
(6,1.20)
(7,2.40)
(8,0.60)
(9,1.00)
(10,0.80)
(11,1.80)
(12,1.60)
(13,0.80)
(14,3.80)
(15,2.60)
(16,1.20)
(17,1.60)
(18,0.80)
};

% Llama
\nextgroupplot[title=Llama]
\addplot+[only marks, mark=square*, color=green!60!black] coordinates {
(1,1.80)
(2,3.60)
(3,1.80)
(4,1.60)
(5,1.40)
(6,1.60)
(7,1.80)
(8,1.20)
(9,1.80)
(10,1.40)
(11,2.20)
(12,1.60)
(13,1.20)
(14,2.40)
(15,3.40)
(16,1.60)
(17,1.60)
(18,1.60)
};

% Claude
\nextgroupplot[title=Claude]
\addplot+[only marks, mark=diamond*, color=purple] coordinates {
(1,1.00)
(2,3.20)
(3,2.00)
(4,1.00)
(5,1.00)
(6,1.60)
(7,2.00)
(8,1.00)
(9,1.20)
(10,1.00)
(11,2.60)
(12,1.80)
(13,1.00)
(14,3.80)
(15,2.80)
(16,1.20)
(17,1.60)
(18,1.40)
};

% Gemini
\nextgroupplot[title=Gemini]
\addplot+[only marks, mark=star, color=orange] coordinates {
(1,0.90)
(2,5.80)
(3,1.60)
(4,1.00)
(5,0.50)
(6,0.80)
(7,2.10)
(8,1.00)
(9,1.00)
(10,1.00)
(11,2.20)
(12,1.50)
(13,1.00)
(14,3.60)
(15,3.10)
(16,1.20)
(17,1.20)
(18,1.00)
};

% Humans
\nextgroupplot[title=Humans Avg]
\addplot+[only marks, mark=*, color=black] coordinates {
(1,3.83)
(2,4.33)
(3,3.11)
(4,3.22)
(5,2.89)
(6,2.56)
(7,3.39)
(8,2.67)
(9,2.89)
(10,2.94)
(11,3.17)
(12,3.06)
(13,2.72)
(14,3.11)
(15,3.28)
(16,3.22)
(17,3.06)
(18,2.78)
};

\end{groupplot}
\end{tikzpicture}
\caption{Scatter plots of risk scores assigned to each question (Q1--Q18) by  LLMs with other professional levels (Non-Senior/Specialist) human experts.}
\label{fig:scatter-llms-humans-no-sen-spec}
\end{figure}
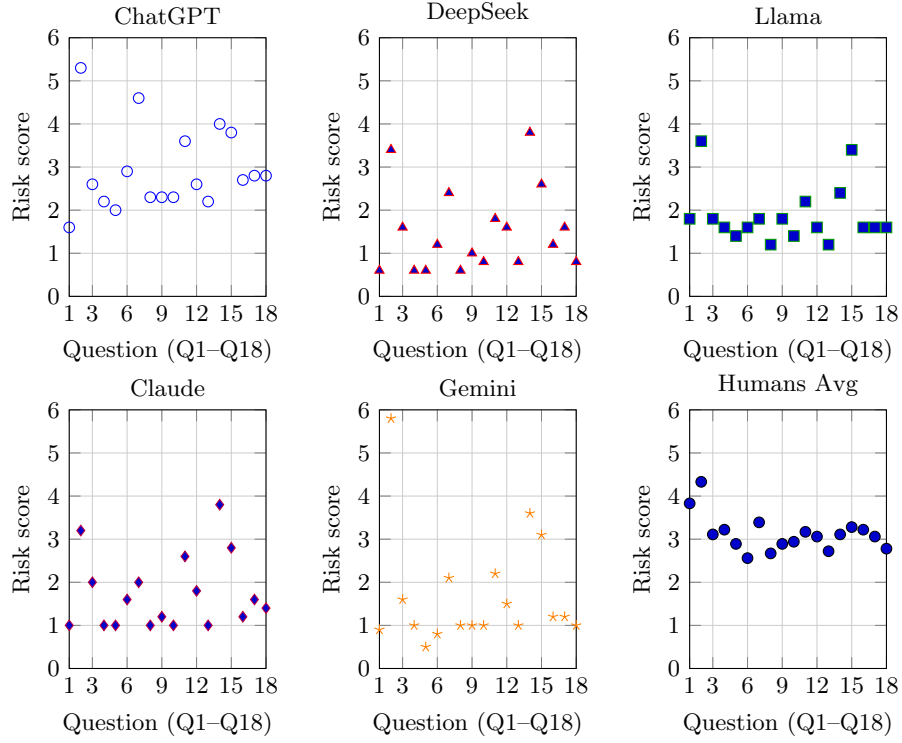

Table \ref{tab:table-average-otherlevel} shows the average risk assigned to each question by each participant across the five models vs. other professional levels surveyed. Figure \ref{fig:scatter-llms-humans-no-sen-spec} presents the scatter plots for each model. It is possible to see that the overall average risk attributed by this group of professionals is higher than that attributed by the five models for all questions, but lower than that of Senior and Specialist participants. In addition to the average risk of \textbf{3.12}, a smaller standard deviation of \textbf{2.41} is observed compared to the group containing all participants and the same group containing Senior and Specialist, which shows the best consensus among these two groups of participants. 
%%%%%%%%%%%%%%%%%%%%%%%%%%%%%%%%%%%%%%%%%%%%%%%%%%%%%%%%%
%Tabela 12 - tabela sumário geral
%%%%%%%%%%%%%%%%%%%%%%%%%%%%%%%%%%%%%%%%%%%%%%%%%%%%%%%%%
\begin{table}[!htbp]
\centering
\caption{Average risk ratings assigned by LLM Models vs other professional levels (Non-Senior/Specialist) human experts for each scenario question.}
\label{tab:table-average-otherlevel}
\setlength{\tabcolsep}{5pt}
\begin{tabular*}{\linewidth}{@{\extracolsep{\fill}} l c c c c c !{\vrule width 0.3pt} c c }
\toprule
\textbf{Q} & \textbf{ChatGPT} & \textbf{DeepSeek} & \textbf{Llama} & \textbf{Claude} & \textbf{Gemini} & \textbf{Human Mean} & \textbf{SD} \\
\midrule
Q1  & 1.60 & 0.60 & 1.80 & 1.00 & 0.90 & 3.83 & 3.22 \\
Q2  & 5.30 & 3.40 & 3.60 & 3.20 & 5.80 & 4.33 & 2.38 \\
Q3  & 2.60 & 1.60 & 1.80 & 2.00 & 1.60 & 3.11 & 2.35 \\
Q4  & 2.20 & 0.60 & 1.60 & 1.00 & 1.00 & 3.22 & 2.88 \\
Q5  & 2.00 & 0.60 & 1.40 & 1.00 & 0.50 & 2.89 & 2.78 \\
Q6  & 2.90 & 1.20 & 1.60 & 1.60 & 0.80 & 2.56 & 1.98 \\
Q7  & 4.60 & 2.40 & 1.80 & 2.00 & 2.10 & 3.39 & 2.40 \\
Q8  & 2.30 & 0.60 & 1.20 & 1.00 & 1.00 & 2.67 & 1.68 \\
Q9  & 2.30 & 1.00 & 1.80 & 1.20 & 1.00 & 2.89 & 2.45 \\
Q10 & 2.30 & 0.80 & 1.40 & 1.00 & 1.00 & 2.94 & 2.34 \\
Q11 & 3.60 & 1.80 & 2.20 & 2.60 & 2.20 & 3.17 & 2.55 \\
Q12 & 2.60 & 1.60 & 1.60 & 1.80 & 1.50 & 3.06 & 2.51 \\
Q13 & 2.20 & 0.80 & 1.20 & 1.00 & 1.00 & 2.72 & 2.11 \\
Q14 & 4.00 & 3.80 & 2.40 & 3.80 & 3.60 & 3.11 & 2.14 \\
Q15 & 3.80 & 2.60 & 3.40 & 2.80 & 3.10 & 3.28 & 2.30 \\
Q16 & 2.70 & 1.20 & 1.60 & 1.20 & 1.20 & 3.22 & 2.71 \\
Q17 & 2.80 & 1.60 & 1.60 & 1.60 & 1.20 & 3.06 & 2.39 \\
Q18 & 2.80 & 0.80 & 1.60 & 1.40 & 1.00 & 2.78 & 2.21 \\

\midrule
\textbf{Total} & \textbf{2.92} & \textbf{1.50} & \textbf{1.87} & \textbf{1.73} & \textbf{1.69} & \textbf{3.12} & \textbf{2.41} \\
\bottomrule
\end{tabular*}
\end{table}

Second, we wanted to know the correlation coefficient between the five models and human experts. In practical terms, there is a direct correlation between the risk ratings assigned by the five models and those designated by human experts. That is, do the risk ratings move in the same direction? To better understand this, we use Pearson's correlation coefficient to measure the strength of the linear relationship between two variables. If there is a strong linear relationship, the correlation coefficient is close to 1 or $-1$; 0 indicates no linear relationship.  

%%%%%%%%%%%%%%%%%%%%%%%%%%%%%%%%%%%%%%%%%%%%%%%%%%%%%%%%%%%%%
% Fig 6 - Gráfico de dispersão RQ2.1 All models vs Other
%%%%%%%%%%%%%%%%%%%%%%%%%%%%%%%%%%%%%%%%%%%%%%%%%%%%%%%%%%%%%

The mean Pearson correlation of \textbf{0.60} shows that while both continue to attribute risk in the same direction, the relationship is weak to modestly positive. Furthermore, the five models continue to rank risk below this group of humans and all others, as noted earlier. (if humans score a particular scenario higher, the five models tend to do so, too—but not as high in absolute terms), as shown in Table \ref{tab:pearson-corr-others-level}.

%%%%%%%%%%%%%%%%%%%%%%%%%%%%%%%%%%%%%%%%%%%%%%%%%%%%%%%%%%%%%
% Tabela 13 Sumarios de coeficientes Pearson All other levels
%%%%%%%%%%%%%%%%%%%%%%%%%%%%%%%%%%%%%%%%%%%%%%%%%%%%%%%%%%%%%
\begin{table}[!htbp]
\centering
\caption{Pearson correlation coefficients between other professional levels (Non-Senior/Specialist) humans experts and LLM models}
\label{tab:pearson-corr-others-level}
\begin{tabular}{lcccccc}
\toprule
& \textbf{LLMs Average} & \textbf{ChatGPT} & \textbf{DeepSeek} & \textbf{Llama} & \textbf{Claude} & \textbf{Gemini} \\
\midrule
Humans & 0.603 & 0.515 & 0.497 & 0.688 & 0.429 & 0.698 \\
\bottomrule
\end{tabular}
\end{table}

Third, we conducted statistical tests (see \ref{tab:paired-t-estothers-level}) to evaluate the significance of differences using paired-samples t-tests for the five models and human experts. For this group of humans, the data again yield an extremely small p-value \textbf{(all $p < 0.001$)}, so we can confidently reject the null hypothesis and assume the alternative hypothesis. There is a significant difference in the average risk ratings between the two groups.  

%%%%%%%%%%%%%%%%%%%%%%%%%%%%%%%%%%%%%%%%%%%%%%%%%%%%%%%%%%%%%
% Tabela 14 Testes pareados, T-test
%%%%%%%%%%%%%%%%%%%%%%%%%%%%%%%%%%%%%%%%%%%%%%%%%%%%%%%%%%%%%
\begin{table}[!htbp]
\centering
\caption{Paired t-test results comparing other professional levels (Non-Senior/Specialist) humans experts with LLM models}
\label{tab:paired-t-estothers-level}
\begin{tabular}{lcc}
\toprule
\textbf{Model} & \textbf{t Stat} & \textbf{p-value (two-tail)} \\
\midrule
LLMs Average & 6.86 & 2.76$\times 10^{-6}$ \\
GPT          & 1.02 & 3.19$\times 10^{-1}$ \\
DeepSeek     & 8.11 & 3.00$\times 10^{-7}$ \\
Llama        & 11.01 & 3.66$\times 10^{-9}$ \\
Claude       & 7.62 & 6.93$\times 10^{-7}$ \\
Gemini       & 5.72  & 2.48$\times 10^{-5}$ \\
\bottomrule
\end{tabular}
\end{table}

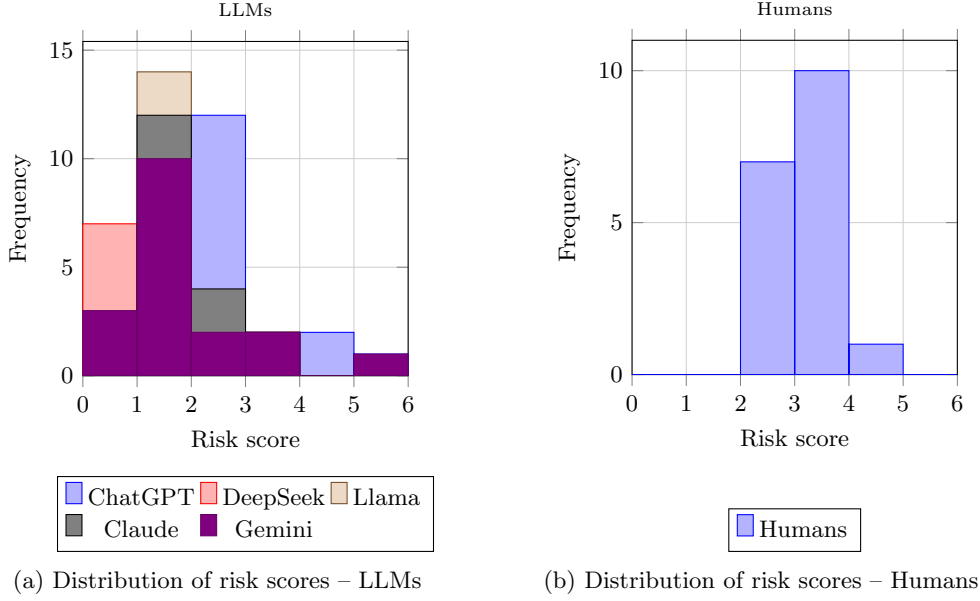
\begin{figure}[!htbp]
\centering
%%%%%%%%%%%%%%%%%%%%%%%%%%%%%%%%%%%%%%%%%%%%%%%%%%%%%%%%%%%%%
% Subfigure Historam LLM
%%%%%%%%%%%%%%%%%%%%%%%%%%%%%%%%%%%%%%%%%%%%%%%%%%%%%%%%%%%%%
\begin{subfigure}[t]{0.45\textwidth}
\centering
\begin{tikzpicture}
\begin{axis}[
  ybar,
  bar width=3pt,
  width=\linewidth,
  height=6cm,
  xlabel={Risk score},
  ylabel={Frequency},
  ymin=0,
  xmin=0, xmax=6,
  xtick={0,1,2,3,4,5,6},
  title={LLMs},
  legend style={
    at={(0.5,-0.3)},
    anchor=north,
    legend columns=3,
    font=\scriptsize}
]
% GPT
\addplot+[
  hist={bins=6, data min=0, data max=6}
] table [y index=0, row sep=\\] {data\\
1.60\\ 5.30\\ 2.60\\ 2.20\\ 2.00\\ 2.90\\ 4.60\\ 2.30\\ 2.30\\ 2.30\\ 3.60\\ 2.60\\ 2.20\\ 4.00\\ 3.80\\ 2.70\\ 2.80\\ 2.80\\};
\addlegendentry{ChatGPT}

% DeepSeek
\addplot+[
  hist={bins=6, data min=0, data max=6}
] table [y index=0, row sep=\\] {data\\
0.60\\ 3.40\\ 1.60\\ 0.60\\ 0.60\\ 1.20\\ 2.40\\ 0.60\\ 1.00\\ 0.80\\ 1.80\\ 1.60\\ 0.80\\ 3.80\\ 2.60\\ 1.20\\ 1.60\\ 0.80\\};
\addlegendentry{DeepSeek}

% Llama
\addplot+[
  hist={bins=6, data min=0, data max=6}
] table [y index=0, row sep=\\] {data\\
1.80\\ 3.60\\ 1.80\\ 1.60\\ 1.40\\ 1.60\\ 1.80\\ 1.20\\ 1.80\\ 1.40\\ 2.20\\ 1.60\\ 1.20\\ 2.40\\ 3.40\\ 1.60\\ 1.60\\ 1.60\\};
\addlegendentry{Llama}

% Claude
\addplot+[
  hist={bins=6, data min=0, data max=6}
] table [y index=0, row sep=\\] {data\\
1.00\\ 3.20\\ 2.00\\ 1.00\\ 1.00\\ 1.60\\ 2.00\\ 1.00\\ 1.20\\ 1.00\\ 2.60\\ 1.80\\ 1.00\\ 3.80\\ 2.80\\ 1.20\\ 1.60\\ 1.40\\};
\addlegendentry{Claude}

% Gemini
\addplot+[
  hist={bins=6, data min=0, data max=6}
] table [y index=0, row sep=\\] {data\\
0.90\\ 5.80\\ 1.60\\ 1.00\\ 0.50\\ 0.80\\ 2.10\\ 1.00\\ 1.00\\ 1.00\\ 2.20\\ 1.50\\ 1.00\\ 3.60\\ 3.10\\ 1.20\\ 1.20\\ 1.00\\ };
\addlegendentry{Gemini}

\end{axis}
\end{tikzpicture}
\caption{Distribution of risk scores – LLMs}
\end{subfigure}%
\hfill
%%%%%%%%%%%%%%%%%%%%%%%%%%%%%%%%%%%%%%%%%%%%%%%%%%%%%%%%%%%%%
% Subfigure Historam Humans
%%%%%%%%%%%%%%%%%%%%%%%%%%%%%%%%%%%%%%%%%%%%%%%%%%%%%%%%%%%%%
\begin{subfigure}[t]{0.45\textwidth}
\centering
\begin{tikzpicture}
\begin{axis}[
  ybar,
  bar width=12pt,
  width=\linewidth,
  height=6cm,
  xlabel={Risk score},
  ylabel={Frequency},
  ymin=0,
  xmin=0, xmax=6,
  xtick={0,1,2,3,4,5,6},
  title={Humans},
  legend style={
    at={(0.5,-0.4)},
    anchor=north,
    legend columns=3,
    font=\scriptsize}
]
\addplot+[
  hist={bins=6, data min=0, data max=6}
] table [y index=0, row sep=\\] {data\\
3.83\\ 4.33\\ 3.11\\ 3.22\\ 2.89\\ 2.56\\ 3.39\\ 2.67\\ 2.89\\ 2.94\\
3.17\\ 3.06\\ 2.72\\ 3.11\\ 3.28\\ 3.22\\ 3.06\\ 2.78\\};
\addlegendentry{Humans}
\end{axis}
\end{tikzpicture}
\caption{Distribution of risk scores – Humans}
\end{subfigure}

\caption{Comparison of distributions of risk scores assigned by LLMs and Human Experts (Q1–Q18).}
\label{fig:hist-llm-vs-human-olevels}
\end{figure}

Finally, the histograms (Figure \ref{fig:hist-llm-vs-human-olevels}) show the distribution of risk ratings assigned by the five models versus those given by human participants. This histogram shows the distribution of risk ratings assigned by the five models compared with those provided by this group of participants. Generally, this group of human participants tends to rate around \textbf{2.6} to \textbf{3.1}, with fewer ratings above \textbf{3.7}, indicating a moderate overall view of risk. These charts highlight how models continue to underestimate risk relative to human respondents' broader, more varied assessments. 

%%%%%%%%%%%%%%%%%%%%%%%%%%%%%%%%%%%%%%%%%%%%%%%%%%%%%%%%%%%%%
% Fig 8 Histrogram of LLM Senior e Spec
%%%%%%%%%%%%%%%%%%%%%%%%%%%%%%%%%%%%%%%%%%%%%%%%%%%%%%%%%%%%%

\subsection{Discussion and limitations}
\label{res:dis}

To enhance clarity and facilitate interpretation, Table~\ref{tab:table-summary-comparative} presents a comparative summary of the main findings from our study. It contrasts the risk assessments produced by the five language models with those of the different groups of human participants, including average risk scores, correlation levels, and statistical significance measures. This summary provides a consolidated view of the key divergences and patterns observed throughout the analysis.

\begin{table}[!htbp]
\centering
\caption{Summary of comparative risk assessment results: The Five Models vs Human Experts}
\label{tab:table-summary-comparative}
\setlength{\tabcolsep}{8pt}
\begin{tabular}{lcccc}
\toprule
\textbf{Group} & \textbf{LLMs Avg} & \textbf{Humans Mean} & \textbf{r} & \textbf{p}\\
\midrule
All Participants       & 1.94 & 3.60 & 0.814 & 3.14$\times 10^{-10}$  \\
Senior \& Specialists  & 1.94 & 3.87 & 0.819 & 1.43$\times 10^{-11}$  \\
Other Levels           & 1.94 & 3.12 & 0.603 & 2.76$\times 10^{-6}$  \\
\bottomrule
\end{tabular}
\end{table}

The comparative results across all three groups indicate that the five models consistently underestimate cybersecurity risks relative to human assessments. This discrepancy was more pronounced among highly experienced professionals (Senior/Specialist), suggesting that domain expertise significantly influences risk perception. The moderate to strong Pearson correlations indicate that the five models align with human trends to some degree, but not to the same extent.

Key lessons learned include:
\begin{itemize}
    \item LLMs follow general human trends, but fail to match expert-level sensitivity to nuanced risks;
    \item Professional background matters: More experienced individuals tend to rate risk higher and more consistently;
    \item Human oversight is crucial, especially in high-stakes decision-making involving incomplete or ambiguous security information.
\end{itemize}

These findings suggest that LLMs should not yet replace human experts in cybersecurity risk assessments, especially for critical decisions. Instead, they may serve as decision-support tools—providing first-pass assessments that require validation. Organizations might benefit from hybrid models in which LLMs help standardize inputs or accelerate assessments, while final judgments remain with qualified professionals.

Furthermore, AI-based assessments could be more reliable in external threat contexts, where risk is more clearly defined and structured. However, internal process evaluations (such as asset inventory or policy adherence) revealed the most significant discrepancies and require deeper contextual interpretation.

While the study reveals important trends and insights, some limitations should be acknowledged:

\begin{itemize}
    \item The analysis focused on only five LLM models and may not reflect performance across other models or future versions.
    \item The participant sample, though diverse, was limited to 50 individuals, which may impact generalizability.
    \item Risk assessment was based on simulated scenarios, which, although grounded in real practices, may not fully reflect operational complexity.
    \item LLMs were tested without enhanced prompt engineering or additional context that might improve performance.
\end{itemize}

Future work should explore prompt optimization, model fine-tuning, and scenario complexity scaling to evaluate how LLM performance can be improved in real-world cybersecurity workflows.

\section{Conclusion and future work}
The results of this study demonstrate a systematic difference between risk assessments made by human experts and those generated by language models. The models generally assign lower risk levels than the experts, suggesting an underestimation of risk in several situations. This discrepancy was statistically significant, suggesting the model may exhibit a conservative bias, particularly when humans perceive high risk. Therefore, AI should be considered a support tool, not a substitute for human expertise.

In addition, we observed that participants' experience influences risk assignment: more experienced and specialized professionals tend to assign higher risk ratings than those with fewer years of experience. This factor reinforces the importance of professional background in risk assessment and suggests that AI models may need adjustments to better capture the nuances of human perception in cybersecurity contexts.

Another relevant finding was the smaller difference between human and LLM risk assessments regarding external threats (such as third-party risks). This may indicate that models better understand external threats than internal ones. On the other hand, the most significant deviations occurred in issues such as asset inventory, where human experts expressed greater concern about partial or incomplete coverage, while the model minimized this risk.

Based on these findings, we propose several directions for future work:

\begin{itemize}
    \item Expansion of the Dataset: Conduct new experiments with a more extensive and more diverse sample of professionals, including different industry sectors, to validate and refine the findings.
    \item Advanced Contextualization: Explore prompt engineering techniques to provide more details to the model, ensuring it better understands the context of the questions and can offer assessments closer to those of human experts.
    \item Hybrid Use of AI and Humans: Future research should explore hybrid approaches in which AI systems act as intelligent assistants to experts, performing preliminary tasks such as suggesting initial classifications, identifying patterns, or highlighting less visible risks.
    \item Evaluation of Other AI Models: Compare the performance of different LLMs in cybersecurity risk assessment, exploring which architectures best suit this task and whether models specifically trained for information security can offer better performance.
\end{itemize}

Through these initiatives, our goal is to advance the use of AI in cybersecurity risk assessment, making the technology a strong, reliable ally in supporting security decision-making within organizations.

\section*{Declarations}\label{declarations}

\subsubsection*{Funding}\label{funding}

The authors thank EVA Cybersec for financial support.

\subsubsection*{Conflict of interest}\label{conflict_interest}

The authors have no competing interests or other interests that might be perceived to influence the results or discussion reported in this paper.

\subsubsection*{Ethics approval}\label{ethics_approval}

This manuscript adheres to the principles and policies of authorship ethics. This study involved the collection of anonymized opinions from voluntary participants. According to Brazilian regulations for research involving human subjects—specifically the Resolution CNS nº 510/2016, which governs ethical oversight for studies in the humanities and social sciences—research activities that gather anonymous opinion data without any possibility of identifying participants are exempt from review by an Institutional Review Board (IRB)/ethics committee. Therefore, no ethics committee approval was required for this study.

\subsubsection*{Consent to participate and publication}\label{consent_participate}

All authors read and approved the final manuscript for publication via the subscription publishing route.

\subsubsection*{Availability of data and materials}\label{availability_data_materials}

All materials used in this manuscript are public, and no permission is required. The results and data in this manuscript have not been published elsewhere.

\subsubsection*{Code availability}\label{code_availability}

All materials used in this manuscript are public, and no permission is required. Additional materials for this article will be available upon request to authors.

\subsubsection*{Authors' contributions}\label{authors_contributions}

The authors contributed equally to this work. 

\bibliography{sn-bibliography}% common bib file

\end{document}